\documentclass[lettersize,journal]{IEEEtran}

\usepackage{amsmath,amsfonts,amssymb}
\usepackage{bm}
\usepackage{textcomp}
\usepackage{stfloats}
\usepackage{url}
\usepackage{verbatim}
\usepackage{graphicx}
\usepackage[shortlabels]{enumitem}
\usepackage{needspace}
\usepackage{balance}
\usepackage{comment}
\usepackage[table]{xcolor}
\usepackage{wasysym}
\usepackage{multirow}
\usepackage[numbers,sort&compress]{natbib}

\usepackage{blindtext}
\usepackage{wrapfig}
\usepackage{booktabs}
\usepackage{tcolorbox}
\tcbuselibrary{breakable}
\usepackage{soul}
\usepackage{colortbl}
\usepackage{listings}
\usepackage{tabularx}
\usepackage{adjustbox}
\usepackage{seqsplit}
\usepackage{xspace}
\usepackage{hyperref}
\usepackage[hyphenbreaks]{breakurl}
\usepackage[ruled,vlined,linesnumbered]{algorithm2e}
\usepackage[caption=false,font=footnotesize]{subfig}

\tcbuselibrary{skins,breakable}

\definecolor{promptgray}{gray}{0.9}
\definecolor{paraorange}{RGB}{59,144,174}
\definecolor{codegreen}{rgb}{0,0.5,0}
\definecolor{framecolor}{rgb}{0.8,0.8,0.8}
\definecolor{stringcolor}{rgb}{0.58,0,0.82}
\definecolor{functioncolor}{RGB}{117,89,49}
\definecolor{codered}{rgb}{0.8,0,0}

\newsavebox{\sourcebox}
\newsavebox{\ghidrabox}
\newsavebox{\idiomsbox}
\newsavebox{\hexbox}
\newsavebox{\gtbox}
\newsavebox{\idiomsboxB}
\newsavebox{\ReSymbox}
\newsavebox{\lohrebox}

\lstdefinestyle{CWithLines}{
    language=C,
    numbers=left,
    numberstyle=\tiny\color{black},
    basicstyle=\ttfamily\scriptsize\color{black},
    keywordstyle=\color{black},
    commentstyle=\color{codegreen},
    stringstyle=\color{stringcolor},
    morekeywords=[1]{__int64, unsigned, signed, int, void, ulonglong, longlong, undefined8, _QWORD, __fastcall},
    keywordstyle=[1]\color{blue},
    morekeywords=[2]{sub_140001100, hash_find_index, tree_find, FUN_140001100, hash_find_index,hash_make_key,gap_get,cmp_item,hash_next_index,tree_find,compare,tree_get,cmp,tree_next,sub_140001100,sub_1400010CC,sub_1400010E8,FUN_140001100,FUN_1400010cc,FUN_1400010e8},
    keywordstyle=[2]\color{functioncolor},
    morekeywords=[3]{a1, a2, h, item, tree, data, param_1, param_2 },
    keywordstyle=[3]\color{paraorange},
    morekeywords=[4]{struct, if, else, while, for, do, switch, case, break, return, continue, goto, default},
    keywordstyle=[4]\color{black},
    breaklines=true,
    frame=single,
    framerule=1pt,
    rulecolor=\color{framecolor},
    xleftmargin=2em,
    keepspaces=true,
    showstringspaces=false,
    alsoletter={_},
    literate=
    {tree*}{{{\color{black}tree*}}}4
    {&}{{\&}}{1}
}

\lstdefinestyle{CaseDeclaration}{
    style=CWithLines,
    basicstyle=\ttfamily\footnotesize,
    numbers=none,
    xleftmargin=0pt,
    aboveskip=3pt,
    belowskip=3pt,
    emphstyle=\color{codered}\bfseries
}

\lstdefinestyle{CWithLinesF2}{
    language=C,
    numbers=left,
    numberstyle=\tiny\color{black},
    basicstyle=\ttfamily\scriptsize\color{black},
    keywordstyle=\color{black},
    commentstyle=\color{codegreen},
    stringstyle=\color{stringcolor},
    morekeywords=[1]{__int64, unsigned, signed, int, void, ulonglong, longlong, undefined8, _QWORD, __fastcall,DWORD},
    keywordstyle=[1]\color{blue},
    morekeywords=[2]{sub_140001100, hash_find_index, tree_find, FUN_140001100, hash_find_index,hash_make_key,gap_get,cmp_item,hash_next_index,tree_find,compare,tree_get,cmp,tree_next,sub_140001100,sub_1400010CC,sub_1400010E8,FUN_140001100,FUN_1400010cc,FUN_1400010e8,Func,comp,hash_func,cmp_func},
    keywordstyle=[2]\color{functioncolor},
    morekeywords=[3]{a1, a2, h, param_1, param_2 },
    keywordstyle=[3]\color{paraorange},
    morekeywords=[4]{struct, if, else, while, for, do, switch, case, break, return, continue, goto, default},
    keywordstyle=[4]\color{black},
    breaklines=true,
    frame=single,
    framerule=1pt,
    rulecolor=\color{framecolor},
    xleftmargin=2em,
    keepspaces=true,
    showstringspaces=false,
    alsoletter={_},
    literate={&}{{\&}}{1}
}

\lstdefinestyle{CWithLinesNoData}{
    language=C,
    numbers=left,
    numberstyle=\tiny\color{black},
    basicstyle=\ttfamily\scriptsize\color{black},
    keywordstyle=\color{black},
    commentstyle=\color{codegreen},
    stringstyle=\color{stringcolor},
    morekeywords=[1]{__int64, unsigned, signed, int, void, ulonglong, longlong, undefined8, _QWORD, __fastcall},
    keywordstyle=[1]\color{blue},
    morekeywords=[2]{sub_140001100, hash_find_index, tree_find, FUN_140001100, hash_find_index,hash_make_key,gap_get,cmp_item,hash_next_index,tree_find,compare,tree_get,cmp,tree_next,sub_140001100,sub_1400010CC,sub_1400010E8,FUN_140001100,FUN_1400010cc,FUN_1400010e8},
    keywordstyle=[2]\color{functioncolor},
    morekeywords=[3]{a1, a2, h, item, tree, param_1, param_2 },
    keywordstyle=[3]\color{paraorange},
    morekeywords=[4]{struct, if, else, while, for, do, switch, case, break, return, continue, goto, default},
    keywordstyle=[4]\color{black},
    breaklines=true,
    frame=single,
    framerule=1pt,
    rulecolor=\color{framecolor},
    xleftmargin=2em,
    keepspaces=true,
    showstringspaces=false,
    alsoletter={_},
    literate={&}{{\&}}{1}
}

\begin{document}

\title{LOHRE: Execution-Grounded Type Recovery via LLM-Orchestrated Hybrid Analysis}

\author{Hao Jiang,
        Yujie Yang,
        Tao Li,
        Chenhui Cui,
        Liangfeng Xu,
        Chengjie Liu,
        Weiping Wen,
        Rubing Huang,~\IEEEmembership{Senior Member,~IEEE},
        and Tsong Yueh Chen,~\IEEEmembership{IEEE Fellow,~IEEE}
\thanks{Hao Jiang, Yujie Yang, Liangfeng Xu, Chengjie Liu and Weiping Wen are with Peking University, Beijing 100871, China (e-mail: haojiang@stu.pku.edu.cn; 2501210755@stu.pku.edu.cn; xuliangfeng26@stu.pku.edu.cn; cheng@stu.pku.edu.cn; weipingwen@pku.edu.cn).}%
\thanks{Tao Li is with Beihang University, Beijing 100191, China (e-mail: abelli@buaa.edu.cn).}%
\thanks{Chenhui Cui and Rubing Huang are with Macau University of Science and Technology, Macao SAR 999078, China (e-mail: 3230002105@student.must.edu.mo; rbhuang@must.edu.mo).}%
\thanks{Tsong Yueh Chen is with Swinburne University of Technology, Melbourne 3122, Australia (e-mail: tychen@swin.edu.au).}%
\thanks{Tao Li and Weiping Wen are the corresponding authors.}
}

\maketitle
\begin{abstract}
Documenting stripped binary functions for security analysis requires recovering function interfaces and user-defined type (UDT) layouts from machine code.
Existing large language model (LLM)-based methods face two limitations: they lack execution-based validation to detect inferred UDTs inconsistent with actual memory layouts, and may overlook evidence about shared UDTs across functions without direct call relationships.
We present LLM-Orchestrated Hybrid Reverse Engineering (\textsc{LOHRE}), a framework that treats inferred UDTs as falsifiable hypotheses and tests them against program execution.
LOHRE uses a hypothesize-validate-refine loop to test candidate layouts through paired dynamic slice emulation of baseline and mutated inputs.
To reduce the cost of processing lengthy execution histories, the Execution-Depth Heuristic (EDH) compresses paired traces into an execution-depth difference and a summary of the tested field's observed uses.
A cross-function knowledge base aggregates compatible UDT hypotheses and field-level confidence scores to guide testing and hypothesis refinement.
We evaluate LOHRE on RealType at four optimization levels and a Windows remote procedure call (RPC) dataset.
LOHRE achieves an average UDT Structural accuracy of 57.85\% across RealType’s four optimization levels and 35.6\% on Windows RPC, outperforming the evaluated static baselines.
Disabling dynamic validation reduces accuracy to 19.28\% on RealType and 9.2\% on Windows RPC.
Removing the cross-function knowledge base reduces accuracy to 41.03\% on RealType and 10.0\% on Windows RPC.
Replacing EDH's compressed feedback with full execution traces reduces recovery accuracy while increasing token use by 3.7 times and runtime by 3.0 times.
These results indicate that execution-based validation, compact feedback, and cross-function evidence contribute to accurate and efficient UDT recovery.
Supplementary materials are available at \url{https://anonymous.4open.science/r/LOHRE-230C}.
\end{abstract}



\begin{lrbox}{\sourcebox}
    \begin{minipage}[t]{0.5\textwidth}
    \begin{lstlisting}[style=CWithLinesNoData]
int hash_find_index(struct hash* h, void* item) {
 void* cnx; int cnt = 0;
 int index = h->hash_make_key(item, h->hash_size);
 cnx = gap_get(h->data, index);
 while (cnx != NULL) {
  if (cnt++ > h->hash_size) return -1;
  if (h->cmp_item(cnx, item) == 0) break;
  index = hash_next_index(h, index);
  cnx = gap_get(h->data,index);
 }
 if (cnx == NULL) return -1;
 return index;
}
\end{lstlisting}
\end{minipage}
\end{lrbox}

\begin{lrbox}{\ghidrabox}
        \begin{minipage}[t]{0.48\textwidth}
        \begin{lstlisting}[style=CWithLines]
ulonglong Ordinal_3(int *param_1,undefined8 param_2) {
 int iVar1; ulonglong uVar2; longlong lVar3; int iVar4;
 uVar2 = (**(code**)(param_1 + 4))(param_2,*param_1);
 iVar4 = 0;
 while( true ) {
  lVar3 = FUN_180001040(*(int **)(param_1 + 2),(int)uVar2);
  if ((lVar3 == 0) || (*param_1 < iVar4)) break;
  iVar1 = (**(code **)(param_1 + 6))(lVar3,param_2);
  if (iVar1 == 0) {
   if (lVar3 != 0) {return uVar2 & 0xffffffff;}
   return 0xffffffff;
  }
  uVar2 = FUN_18000105c(param_1,(int)uVar2);
  iVar4 = iVar4 + 1;
 }
 return 0xffffffff;
}
\end{lstlisting}
\end{minipage}
\end{lrbox}

\begin{lrbox}{\hexbox}
        \begin{minipage}[t]{0.5\textwidth}
        \begin{lstlisting}[style=CWithLines]
__int64 __fastcall test_3(unsigned int *a1, __int64 a2) {
 unsigned int v4; __int64 v5;  __int64 v6; signed int v7;
 signed int v8; unsigned int v9; __int64 v10; __int64 v11;
 v4 = (*((__int64 (__fastcall **)(__int64,_QWORD))a1+2))(a2,*a1);
 v7 = 0;
 while ( 1 ) {
  v9 = v4;
  v10 = sub_180001040(*((_QWORD *)a1 + 1), v4, v5, v6);
  v11 = v10;
  if ( !v10 ) return 0xFFFFFFFFLL;
  v8 = v7++; if ( v8 > (int)*a1 ) return 0xFFFFFFFFLL;
  if (!(*((unsigned int (__fastcall **)(__int64, __int64)) a1 + 3))(v10, a2)) break;
  v4 = sub_18000105C(a1, v9);
 } if ( !v11 ) return (unsigned int)-1;
 return v9;
}
\end{lstlisting}
\end{minipage}
\end{lrbox}

\begin{lrbox}{\idiomsbox}
        \begin{minipage}[t]{0.48\textwidth}
        \begin{lstlisting}[style=CWithLines]
int tree_find(struct tree* tree, void* data) {
 int i; int j; void *node;
 i = tree->compare(data, tree->root);
 for (j = 0; ; j++) {
  node = tree_get(tree->nodes, i);
  if (!node) return -1;
  if (j > tree->root) return -1;
  if (!tree->cmp(node, data)) break;
  i = tree_next(tree, i);
 }
 if (!node) return -1;
 return i;
}
\end{lstlisting}
\end{minipage}
\end{lrbox}

\begin{lrbox}{\gtbox}
    \begin{minipage}[t]{0.428\textwidth}
        \begin{lstlisting}[style=CWithLinesF2]
struct gap_array { int len; void** Array;};
struct hash {
 int hash_size; int item_cnt;
 struct gap_array* data;
 int (*hash_make_key)(void* item, int hash_size);
 int (*cmp_item)(void* item1, void* item2);
};
\end{lstlisting}
    \end{minipage}
\end{lrbox}

\begin{lrbox}{\idiomsboxB}
    \begin{minipage}[t]{0.428\textwidth}
        \begin{lstlisting}[style=CWithLinesF2]
struct tree {
 int root; void *nodes;
 int (*compare)(void *, void *);
 int (*cmp)(void *, void *);
};
struct tree_node {
 int parent; int left; int right;
};
struct tree_data {
 int data; int parent; int left; int right;
};
\end{lstlisting}
    \end{minipage}
\end{lrbox}

\begin{lrbox}{\ReSymbox}
    \begin{minipage}[t]{0.565\textwidth}
        \begin{lstlisting}[style=CWithLinesF2]
struct HashTable{
 unsigned int Slots; unsigned int Count;
 HashFunctions* Func; HashFunctions* comp;
}
\end{lstlisting}
    \end{minipage}
\end{lrbox}

\begin{lrbox}{\lohrebox}
    \begin{minipage}[t]{0.565\textwidth}
    \begin{lstlisting}[style=CWithLinesF2]
struct HashTable {
 DWORD capacity; // Number of slots
 DWORD size; // Number of elements
 Container* container; // Pointer to container
 hash_func_t hash_func; // Pointer to hash func
 compare_func_t cmp_func; // Pointer to cmp func
};
struct Container {
 DWORD capacity; // Number of elements in the array
 DWORD padding; // Alignment padding(4 bytes)
 void** elements; // Pointer to array of elements
};
typedef unsigned int (__fastcall* hash_func_t)(int* key, DWORD capacity);
typedef int (__fastcall* compare_func_t)(void* element, int* key);
\end{lstlisting}
\end{minipage}
\end{lrbox}


\begin{figure*}[!t]
    \centering
    \begin{minipage}[t]{0.5\textwidth}
        \subfloat[Source Code\label{fig:sourceCode}]{\resizebox{\linewidth}{!}{\usebox{\sourcebox}}}
    \end{minipage}
    \hspace{0.01\textwidth}
    \begin{minipage}[t]{0.47\textwidth}
        \subfloat[Decompiled by Idioms\label{fig:idiomssource}]{\resizebox{\linewidth}{!}{\usebox{\idiomsbox}}}
    \end{minipage}

    \begin{minipage}[t]{0.5\textwidth}
        \subfloat[Decompiled by Hex-Rays (IDA Pro)\label{fig:idaCode}]{\resizebox{\linewidth}{!}{\usebox{\hexbox}}}
    \end{minipage}
    \hspace{0.01\textwidth}
    \begin{minipage}[t]{0.47\textwidth}
        \subfloat[Decompiled by Ghidra\label{fig:ghidrasource}]{\resizebox{\linewidth}{!}{\usebox{\ghidrabox}}}
    \end{minipage}
    \caption{Comparison of source code with output from three reverse-engineering tools.}
    \label{fig:code_comparison}
\end{figure*}


\section{Introduction}
\label{sec:intro}
\IEEEPARstart{S}{ecurity} analysis of closed-source software requires recovering high-level information from stripped machine code~\cite{shoshitaishvili2016sok}.
A central part of this task is documenting stripped binary functions by recovering their interfaces and user-defined type (UDT) layouts.
UDTs, such as structs and unions, describe how fields and nested structures are organized in memory.
Without precise UDTs, downstream security tasks such as structure-aware fuzzing have difficulty in constructing valid structured inputs and reaching deeper program states~\cite{RN21}.
However, conventional decompilers such as Ghidra and Hex-Rays expose memory accesses through pointer offsets, often leaving analysts to reconstruct the underlying structures by hand.

Recent large language model (LLM)-based methods~\cite{dramko2025idioms,RN22,tan2024llm4decompile} automate parts of this task, but two limitations remain central to our setting.
First, without execution-based validation, an inferred UDT layout may appear plausible even when it is inconsistent with the program's memory accesses.
Figure~\ref{fig:code_comparison} illustrates this hallucination risk: Idioms~\cite{dramko2025idioms} mistakes a hash table for a tree.
Second, evidence about a shared UDT may be distributed across functions without direct call relationships, each exposing only part of the layout.
Methods that primarily aggregate evidence along call relationships may therefore overlook some of this UDT information~\cite{dramko2025idioms}.
This problem is especially visible in Windows remote procedure call (RPC) services and may also arise when an object's lifetime spans multiple functions, as in object-oriented systems, plugins, and middleware.

We propose \textbf{L}LM-\textbf{O}rchestrated \textbf{H}ybrid \textbf{R}everse \textbf{E}ngineering (\textsc{LOHRE}), a training-free framework for execution-grounded type recovery.
LOHRE treats LLM-inferred UDT layouts as falsifiable hypotheses and tests candidate layouts through paired dynamic slice emulation.
For each test, it executes the target function with a baseline input constructed from the current layout and a mutated input that substitutes an alternative type of equal byte width for a single field.
Comparing the two runs helps determine which type better matches the field's actual usage in the program.
However, program execution histories are often long and complex, which can degrade LLM reasoning and increase token costs~\cite{modarressi2025nolima,du2025context, gao2026ink,zeng2026loca,kim2026defensive}.
To address this problem, we propose the Execution-Depth Heuristic (EDH), which compresses paired execution traces into a depth difference and a summary of the selected field's observed uses.
The depth difference provides a heuristic signal about execution progress, as inputs constructed from incompatible layouts may fail early checks.
The LLM interprets this signal together with the field's observed uses and relevant code to refine the hypothesis.
To combine scattered evidence, LOHRE maintains a cross-function \emph{knowledge base} (KB) that aggregates UDT hypotheses with compatible field offsets and types, retaining their field-level confidence scores. 
These confidence scores focus later testing on uncertain fields, allowing earlier findings to support further recovery.
If analysis of an execution anomaly contradicts an inherited hypothesis, the affected hypothesis is reopened for refinement.
The recovered types and inferred behavior are then organized into documentation for the target function.

Our main contributions are as follows:

\begin{itemize}
    \item 
    We introduce a hypothesize-validate-refine paradigm that treats LLM-inferred UDT layouts as falsifiable hypotheses tested through paired dynamic slice emulation, with observed program behavior guiding subsequent type refinement.
     
    \item 
    We propose the Execution-Depth Heuristic (EDH), which compresses paired baseline and mutated execution traces into a depth difference and a summary of the tested field's observed uses to guide type refinement.
    
    \item 
    We design a cross-function knowledge base that aggregates compatible UDT hypotheses across functions, including those without direct call relationships, retains field-level confidence scores, and supports hypothesis refinement when conflicting execution evidence emerges.
    
    \item 
    We implement and evaluate LOHRE using the RealType dataset compiled at four optimization levels and a Windows RPC dataset.
    LOHRE achieves UDT Structural accuracy of 57.85\% on RealType and 35.6\% on Windows RPC, outperforming the evaluated static baselines.
    Ablation studies show that dynamic validation, EDH, and the cross-function knowledge base each contribute to LOHRE's recovery performance.
\end{itemize}

\section{Background and Motivation}
\label{sec:background}

\subsection{The Semantic Gap and the UDT Recovery Problem}
\label{sec:gap}

In a stripped binary, source-level identifiers, comments, and type information are generally unavailable, although the executable code preserves the program's behavior~\cite{shoshitaishvili2016sok}.
Binary function documentation helps bridge this gap by explaining a function's interface and behavior.
Such documentation relies on type recovery to identify the types of function parameters and accessed data.
A particularly challenging part of type recovery is UDT recovery, which reconstructs the memory layout of fields and padding.
Without this layout, downstream tasks like structure-aware fuzzing struggle to construct valid inputs.

Current automated tools can only partially recover UDTs.
As Fig.~\ref{fig:code_comparison} illustrates, Hex-Rays (IDA Pro's decompiler) and Ghidra surface only pointer-offset accesses.
LLM-based methods such as Idioms~\cite{dramko2025idioms} produce more semantic output but, without validation, can hallucinate entirely incorrect structures.
Idioms mistakes a hash table with callbacks for a tree in Fig.~\ref{fig:code_comparison}(b) and introduces unsupported node structures in Fig.~\ref{fig:udt_comparison_stacked}(c).
The source in Fig.~\ref{fig:udt_comparison_stacked}(a) contains a \texttt{hash} structure whose \texttt{data} field points to a nested \texttt{gap\_array} structure.
ReSym identifies the top-level hash-table structure but misses the \texttt{data} field, while our LOHRE is able to recover both the top-level \texttt{HashTable} structure and its nested \texttt{Container} structure.

\begin{figure*}[!t]
    \centering
    \begin{minipage}[t]{0.435\textwidth}
        \subfloat[Source\label{fig:gt_hash}]{\resizebox{\linewidth}{!}{\usebox{\gtbox}}}
    \end{minipage}
    \hspace{0.001\textwidth}
    \begin{minipage}[t]{0.54\textwidth}
        \subfloat[ReSym\label{fig:ReSym_hash}]{\resizebox{\linewidth}{!}{\usebox{\ReSymbox}}}
    \end{minipage}

    \begin{minipage}[t]{0.435\textwidth}
        \subfloat[Idioms\label{fig:idioms_hash}]{\resizebox{\linewidth}{!}{\usebox{\idiomsboxB}}}
    \end{minipage}
    \hspace{0.001\textwidth}
    \begin{minipage}[t]{0.54\textwidth}
        \vspace{-7.2mm}
        \subfloat[LOHRE\label{fig:lohre_hash}]{\resizebox{\linewidth}{!}{\usebox{\lohrebox}}}
    \end{minipage}
    \caption{Recovery of nested hash-table structures by ReSym, Idioms, and LOHRE.}
    \label{fig:udt_comparison_stacked}
\end{figure*}

\subsection{Limitations of Static-Only LLM Methods}
\label{sec:limits}

Existing LLM-based methods generally refine decompiler output~\cite{wang2025typeforge,hu2024degpt,sha2025hyres} or perform end-to-end recovery~\cite{tan2024llm4decompile,RN22,dramko2025idioms}.
Two limitations affect UDT recovery: unchecked layout predictions and scattered evidence.

\textbf{Unchecked layout predictions.} Without execution feedback, an inferred layout may remain plausible even when it is inconsistent with the program's memory accesses.
A field's width alone may support several types.
For example, an 8-byte field could hold a pointer or a 64-bit scalar.
The Systematization of Knowledge (SoK) on LLMs for reverse engineering~\cite{hu2025sok} identifies robustness and hallucination mitigation as important but understudied objectives.
Figure~\ref{fig:code_comparison} and Figure~\ref{fig:udt_comparison_stacked} provide a concrete example of this failure mode.

\textbf{Scattered evidence in modular binaries.}
Exported application programming interface (API) functions can access different fields of a shared UDT without calling one another.
Call-graph-based aggregation may miss this shared-data relationship and recover incomplete UDTs~\cite{dramko2025idioms}.
The same problem can arise in object-oriented systems, plugin frameworks, and event-driven middleware when an object's lifetime spans several handlers.

Addressing these limitations requires execution-based validation and cross-function evidence reuse. 
A practical framework must also manage execution feedback efficiently and minimize model preparation costs.

\textbf{Dynamic execution feedback.}
Dynamic execution feedback can help distinguish competing types when execution reaches an operation that accesses the selected field.
For example, observing a field value used as a memory address supports a pointer interpretation, whereas an early return before that use leaves the hypothesis unresolved.
Effective validation therefore requires both targeted inputs and field-specific execution observations.

\textbf{Execution feedback interpretation.}
Dynamic execution, however, can produce long execution traces that increase token costs and obscure evidence relevant to a tested field.
A larger context window does not eliminate this difficulty, as studies report degraded performance even within supported input lengths~\cite{liu2024lost,modarressi2025nolima,du2025context}.
The feedback should therefore retain the field's observed uses and relevant execution outcomes in a compact form.

\textbf{Model preparation.}
Idioms~\cite{dramko2025idioms}, ReSym~\cite{RN22}, and LLM4Decompile~\cite{tan2024llm4decompile} use models prepared through fine-tuning or domain adaptation.
This dependence can require multi-day GPU training and may limit transfer to other architectures or coding styles.
These limitations motivate a framework that can use an existing LLM without task-specific fine-tuning.

\begin{figure*}[!t]
    \centering
    \includegraphics[width=\textwidth]{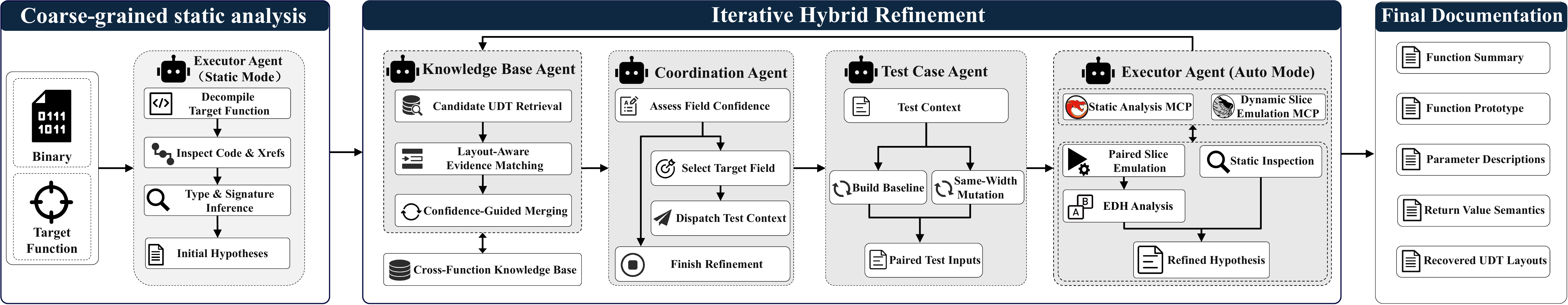}
    \caption{LOHRE's workflow for iterative hypothesis testing and documentation.}
    \label{fig:architecture}
\end{figure*}

\subsection{Motivation: A Hybrid, LLM-Orchestrated Loop}
\label{sec:motivation}

Addressing the two central limitations requires complementary kinds of evidence.
Dynamic execution can invalidate a plausible but incorrect layout, while cross-function analysis can recover UDT information that no single function exposes.
LOHRE combines these two distinct types of evidence in a hypothesize-validate-refine workflow.

LOHRE uses static analysis to propose structural candidates, dynamic emulation to test inputs constructed from them, and EDH to summarize the resulting execution evidence.
For example, a decrease in execution depth may reflect an incompatible type or a change in control flow unrelated to that type.
LOHRE examines the field's observed uses and relevant code to interpret this signal and decide how to refine the hypothesis.
The knowledge base carries compatible hypotheses and their field confidences across functions, so later observations can extend or revise an earlier interpretation.
%

\section{Approach}
\label{sec:approach}

LOHRE implements the loop with four specialized agents: the Coordination Agent, Knowledge Base Agent, Test Case Agent, and Executor Agent (Fig.~\ref{fig:architecture}).
They interact with two external tools through the Model Context Protocol (MCP)~\cite{mcp}: the Static Analysis MCP and the Dynamic Slice Emulation MCP.
This division of responsibilities allows each agent to maintain a compact, task-focused context, thereby curbing the overhead of repeatedly processing accumulated execution histories~\cite{kim2026defensive} and mitigating the reasoning degradation caused by excessively long inputs~\cite{modarressi2025nolima,du2025context}.
The framework's token efficiency is quantified in Section~\ref{sec:rq5}.

Each refinement step uses a concise record linking the selected field and its current hypothesis to the supporting EDH feedback.
Evidence locations are retained so that the Executor can inspect the corresponding code when more detail is needed to support a decision.
The record thus keeps the working context focused while preserving the basis for checking and revising an earlier interpretation.

The supplementary document specifies the four agents' tasks and reasoning rules.
The artifact linked at the end of the abstract provides the prompt catalog, including the full public system texts and shared guidance.

\subsection{Workflow}
\label{sec:workflow}

Analysis of a target function follows the three stages in Fig.~\ref{fig:architecture}.

\textbf{Coarse-grained static analysis.} Given a binary and a target function, LOHRE prepares the analysis environment.
The Executor Agent in Static mode decompiles the function, inspects code and cross-references, and infers preliminary types and the function signature.
These initial hypotheses, together with field-level confidence scores, establish the analysis state for refinement.

\textbf{Iterative hybrid refinement.} The Knowledge Base Agent retrieves candidate UDTs, matches layout evidence, and merges compatible hypotheses using their field confidences.
The Coordination Agent assesses the resulting confidences and either finishes refinement or selects the lowest-confidence target field.
It dispatches a \emph{test context} containing the selected field and the current analysis information needed to construct a test.
The Test Case Agent uses this context to build a baseline input and a same-width mutation, producing paired test inputs in recursively nested \texttt{[type, content, length]} syntax.
The Executor Agent in Auto mode performs paired slice emulation, interprets the EDH feedback, and uses static inspection when further evidence is needed.
The updated analysis state returns to knowledge synthesis for the next coordination step.
Refinement stops when all relevant field confidences reach 0.9 or after 20 paired-test iterations for the function.

\textbf{Final documentation.} The Knowledge Base Agent's report mode formats the latest analysis state into a function summary, function prototype, parameter descriptions, return value semantics, and recovered UDT layouts.
This state includes the latest Executor result when the iteration cap is reached.

\begin{algorithm}[!t]
\caption{Iterative Hybrid Refinement}
\label{alg:coordinator}
\KwIn{Target function $f$, iteration cap $N_{\max}=20$,
      confidence threshold $\tau=0.9$}
\KwOut{Documentation report $R$}
$D \leftarrow \textsc{Executor.analyze}(f, \texttt{Static\_Only})$\;
\For{$i = 1$ \textbf{to} $N_{\max}$}{
    $D \leftarrow \textsc{KnowledgeBase.synthesize}(D)$\;
    $\textit{dec} \leftarrow \textsc{Coordination.assess}(D, \tau)$\;
    \If{$\textit{dec.status} = \text{Over}$}{\textbf{break}}
    $T \leftarrow \textsc{Coordination.dispatch}(D,
                  \textit{dec}.\text{mutate\_field})$\;
    $P \leftarrow \textsc{TestCase.generate}(T)$\;
    $D \leftarrow \textsc{Executor.analyze}(f, \texttt{Auto}, D, P)$\;
}
\Return $\textsc{KnowledgeBase.generate\_report}(f, D)$\;
\end{algorithm}

\subsection{Agents}
\label{sec:agents}

The four agents handle coordination, cross-function knowledge synthesis and report formatting, paired test-input construction, and hybrid analysis, respectively.
Static and Auto are two modes of the same Executor Agent.

\textbf{Coordination Agent.}
Algorithm~\ref{alg:coordinator} summarizes the function analysis shown in Fig.~\ref{fig:architecture}.
The analysis state $D$ contains the parameter template, UDT hypotheses, function summary, field confidences, and available execution feedback.
The relevant confidence entries cover the parameters and UDT members eligible for testing; the selected entry is called the target field in the workflow.
After knowledge synthesis, the Coordination Agent checks whether all relevant confidences reach 0.9.
If all relevant confidences reach 0.9, the Coordination Agent returns \texttt{Over}; otherwise, it selects the lowest-confidence target field and dispatches the associated test context to the Test Case Agent.
The context provides the selected field, current parameter template, UDT hypotheses, function summary, and available execution feedback.
These inputs let test construction follow the current interpretation and the conditions revealed by earlier runs.

\textbf{Knowledge Base Agent.}
During refinement, the Knowledge Base Agent retrieves candidate UDT definitions from the cross-function knowledge base and compares them with the current hypotheses.
It performs layout-aware evidence matching using the structure-aggregation method proposed by HyRES~\cite{sha2025hyres}.
Field names are compared using general text embeddings (GTE)~\cite{li2023towards}, together with checks on offset alignment and field-type consistency.
LOHRE refines HyRES's layout-overlap penalty for unresolved layout regions provisionally represented as padding.
Overlap in these regions is not penalized, allowing later observations to replace the unresolved bytes with supported members.
The agent justifies each merge, rejects inconsistent offsets or field types, and creates a new entry when a hypothesis has no compatible match.

Confidence-guided merging prioritizes higher-confidence field information among compatible hypotheses and preserves the associated scores in the resulting definition.
The Coordination Agent can then focus testing on the least certain fields, reducing repeated work.
If execution analysis identifies an inherited field as the cause of an anomaly, the Executor lowers that field's confidence and retests the hypothesis.
Later functions can thus both extend and correct earlier definitions.
For final documentation, the Knowledge Base Agent formats the latest analysis state into the five components shown in Fig.~\ref{fig:architecture}: function summary, function prototype, parameter descriptions, return value semantics, and recovered UDT layouts.

\textbf{Test Case Agent.}
The Test Case Agent receives the test context dispatched by the Coordination Agent and constructs baseline and mutated inputs in \texttt{[type, content, length]} syntax.
For a scalar, \texttt{content} holds its value.
For a structure pointer, \texttt{["struct\_pointer", [\ldots], p]} contains an ordered list of member triplets, with recursive nesting for structures and arrays.
Here, \texttt{p} is the pointer width in bytes, which is 8 on x86-64.
The pair is designed to reach an operation that can distinguish the candidate types.
The agent identifies the required conditions from the current function summary, input checks, and previous execution feedback.
If reaching the operation depends on other arguments or global settings, the Test Case Agent configures those arguments or settings consistently for both runs.
A compatible input that previously returned normally can serve as a starting point for testing another field.

To form the pair, the agent builds a baseline under the current hypothesis and applies a same-width mutation to the selected field.
The mutation tests a competing type suggested by the field's apparent role while preserving the field's storage width in bytes.
An 8-byte field, for example, may be tested as a 64-bit integer, a double, or a pointer while preserving the surrounding layout.
For a pointer input, the agent describes the pointed-to storage using the current UDT definition and recursively nested member triplets.
The Dynamic Slice Emulation MCP allocates these objects and binds their addresses when preparing each run.
All other arguments, fields, and prepared execution conditions remain consistent across the pair.
In subsequent rounds, revised function summaries and execution feedback guide further input adjustments.

\textbf{Executor Agent.}
In Static mode, exposed as \texttt{Static\_Only} in the interface, the Executor decompiles the target function, inspects code and cross-references, and forms initial UDT and parameter hypotheses.
Its initial field confidences are capped at 0.5.
In Auto mode, exposed as \texttt{Auto}, it runs the paired test inputs through the Dynamic Slice Emulation MCP and interprets the compressed EDH feedback (Section~\ref{sec:edh}).
Static inspection through the Static Analysis MCP supplies further code evidence when needed.
The Executor then updates the UDT interpretation, field confidences, and function summary according to the resulting evidence.
Each confidence score expresses the Executor's current assessment of the supporting evidence, with 0.9 as the acceptance and stopping threshold.
When a definition is reused across functions, its field confidences are retained, allowing refinement to concentrate on less certain fields.

When EDH feedback leaves competing types unresolved, the Executor can request detailed traces of the baseline and mutated runs to locate where their execution diverges.
It then uses static inspection of the corresponding code to explain the divergence and revise the selected field's assessment and type hypothesis.
This analysis may identify an inherited field as the cause of an execution anomaly, even if another field is being tested.
In that case, the Executor lowers the responsible field's confidence and reopens its type hypothesis.
The Executor also uses these findings to revise the function summary when they reveal an input condition or clarify the function's behavior.
For instance, if execution stops before using the selected field, the Executor records the blocking input condition to guide subsequent test construction.

Execution can also help complete an initial layout by revealing accesses within an unknown region of an identified object.
The Executor uses the access offset within that object and the access width to propose a member, preserving the established object size and neighboring members.
The proposed member then enters subsequent tests as a tentative hypothesis, allowing refinement to add members suggested by actual accesses as well as revise existing types.

\textbf{Name inference.}
To describe the roles of recovered parameters and members, the Executor proposes names using static code context and API usage.
Execution evidence helps clarify these roles by showing how the function uses the corresponding values.
The Executor stores the proposed names alongside the type hypotheses.
The Knowledge Base Agent uses these records to assemble the final report.
The evaluation then compares these names with reference identifiers to measure name accuracy separately from type accuracy.

\subsection{MCP Toolset}
\label{sec:mcp}

The Model Context Protocol~\cite{mcp} provides a uniform interface for LLM-tool interaction and allows compatible tool implementations to be substituted.
The Static Analysis MCP exposes decompilation, disassembly, and cross-reference queries so that the Executor can inspect the code under a type hypothesis.
The Dynamic Slice Emulation MCP tests that hypothesis through paired slice emulation with the constructed inputs and collects the execution evidence needed for EDH analysis.

Before execution, the dynamic tool loads the target binary and its dependencies, allocates memory for pointer objects, and places arguments according to the target's calling convention.
To make the comparison meaningful, both runs start from the same initial conditions, including the loaded image's global values and any adjustments needed to reach the selected field.
Their execution states are separate, preventing memory or resource changes in one run from carrying into the other.
During execution, API models or configured replacement functions (stubs) handle external calls, and the tool records the information needed to construct EDH feedback.

Once this environment is prepared, execution begins at the target function's entry and follows its control flow through any called functions.
Execution stops when the target returns, an exception occurs, or the execution budget is reached.
In LOHRE, dynamic slice emulation denotes this execution of the target function and its callees under prepared inputs and isolated runtime state.
Paired slice emulation executes the target once with the baseline input and once with the mutated input under the same prepared conditions.
The emulation tool records basic blocks, external calls, and execution outcomes for each run.
It also records how the selected field's value and values derived from it are used in memory accesses, conditions, computations, and API arguments, supplying the observations that EDH summarizes for hypothesis assessment.

\subsection{Execution-Depth Heuristic}
\label{sec:edh}

\begin{figure}[!t]
    \centering
    \includegraphics[width=0.95\columnwidth]{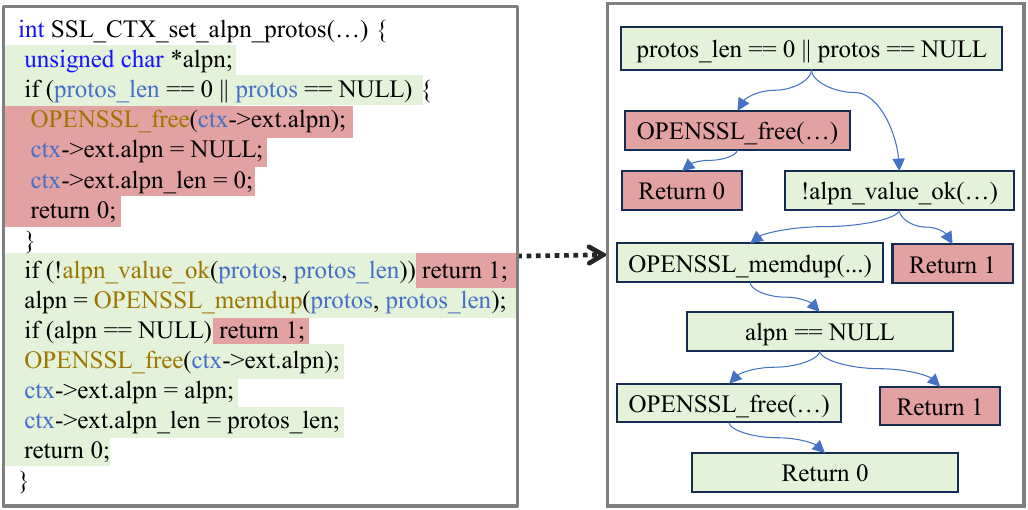}
    \caption{Control flow of OpenSSL's \texttt{SSL\_CTX\_set\_alpn\_protos}.}
    \label{fig:depth_heuristic}
\end{figure}

EDH compresses paired execution traces into a depth difference and a short behavior summary for the selected field.
This feedback keeps type refinement focused and reduces the cost of processing lengthy execution histories.
The depth comparison reflects the input checks that often precede a function's main logic.
An input constructed from an incompatible layout may fail these checks, whereas one constructed from a compatible layout can enable further execution.
Figure~\ref{fig:depth_heuristic} illustrates this pattern in \texttt{SSL\_CTX\_set\_alpn\_protos}, with green marking the main processing path and red marking early-return branches.
To quantify this difference, let $B_M(r)$ be the set containing the basic-block entry addresses executed during run $r$ in the target's binary module $M$.
A basic block is a sequence of instructions with no internal branch.
We define execution depth as
\[
    d(r) = |B_M(r)| + N_{\mathrm{API}}(r),
\]
where $N_{\mathrm{API}}(r)$ counts observed external API calls.
Each block contributes once, including blocks in callees within the same module.
External library internals are outside the block count, while repeated API calls contribute individually.
For paired runs, a positive $\Delta d = d(r_{\mathrm{mutated}})-d(r_{\mathrm{baseline}})$ initially favors the mutation, a negative value favors the baseline, and zero gives no preference from depth.

To relate this comparison to the field hypothesis being tested, EDH summarizes how the target uses the selected value.
The execution tool first marks the value at its input location in a register or memory.
It follows that origin through copies and supported computations, recording derived values and their uses in the target module.
When a write occurs, the destination receives the new value's origin record if the value has been traced to the selected field.
Otherwise, the tool clears the old record so that it no longer attributes the destination's contents to an earlier value.
At external calls, known API declarations associate tracked values with the receiving parameters, distinguishing field values from addresses of field storage.
Because repeated execution can generate redundant observations, the tool merges records with the same runtime instruction address, use category, and descriptive attributes.
A use category describes the value's role, such as addressing memory or supplying an API argument.
Attributes such as access width and argument position distinguish different uses at the same instruction.
Each merged record retains occurrence counts and representative operand values, and matching records are aligned between the runs.
Within this summary, address uses and declared API arguments receive priority because they directly constrain interpretation.
Conditions and computations provide further distinctions between value types.
These records and execution outcomes form the compact behavior summary accompanying the depth difference.

The Executor uses this summary to determine whether the depth preference supports the candidate type.
For example, a four-byte read establishes an access width, whereas using the value as a memory address supports a pointer interpretation.
Passing the selected value to a known integer API parameter can also constrain its interpretation, even when the paired runs have equal depth.
The Executor uses these constraints from normally returning runs to refine the candidate.
If the uses establish only a broad type category, the more precise declaration remains open to further refinement.
When the evidence conflicts or needs further explanation, the recorded code locations guide static inspection and requests for detailed traces.

\subsection{Worked Example: One Iteration}
\label{sec:walkthrough}

To illustrate the loop, we trace one iteration on \texttt{hash\_find\_index} (Fig.~\ref{fig:udt_comparison_stacked}).
The coarse-grained static pass recovers a preliminary \texttt{HashTable} definition with five fields, but the 8-byte field at offset 8 remains uncertain.
The function passes this field's value to \texttt{FUN\_180001040}, the hash table's \texttt{gap\_get} helper.
This observation establishes the field's width but leaves open whether its value is a pointer to a nested structure or a 64-bit scalar.
The Coordination Agent selects this field for the next test.

The Test Case Agent emits paired test inputs using the format described in Section~\ref{sec:agents}.
The baseline encodes the field as \texttt{["struct\_pointer",\allowbreak [["int",1,4], ["int",2,4],\allowbreak ["pointer",arr,8]],8]}, i.e., a pointer to a nested \texttt{Container} that mirrors the \texttt{gap\_array} layout: a 4-byte length, 4 bytes of alignment padding, and an 8-byte array pointer. The mutation preserves the 8-byte width but encodes the field as \verb|["qword",k,8]|, a 64-bit scalar.
The Executor runs both inputs through the Dynamic Slice Emulation MCP.

In the baseline run, \texttt{FUN\_180001040} uses the field as a valid structure pointer, reads \texttt{len} and \texttt{Array}, and returns a live entry.
Control then returns to \texttt{hash\_find\_index}, which continues through comparison and advancement before reaching its success branch.
In the mutated run, the field instead contains a 64-bit scalar.
When \texttt{FUN\_180001040} uses this value as an address, an invalid memory read stops execution at the first dereference.
EDH favors the baseline because it reaches more code, and the behavior summary connects this difference to the tested field: the pointer supports nested-object accesses, whereas using the scalar as an address causes a fault.
These observations support the \texttt{struct\_pointer}-to-\texttt{Container} interpretation, so the Executor raises the field's confidence past the threshold.
The Knowledge Base Agent then incorporates the hypothesis.
This decision follows from paired EDH feedback without consulting per-function ground truth.

\section{Evaluation}
\label{sec:evaluation}

LOHRE organizes the recovered types and inferred behavior into a report describing the function's purpose, interface, and UDT layouts.
We assess recovery accuracy using the exact-match metrics established by Idioms~\cite{dramko2025idioms}, allowing direct comparison with existing methods.
A complementary user study examines whether practitioners find the resulting documentation accurate and useful beyond exact textual matches.
Five research questions guide the evaluation:

\begin{description}
\item[\textbf{RQ1}] How does LOHRE compare with the evaluated baselines on Variable Name, Variable Type, UDT Nominal, and UDT Structural accuracy?
\item[\textbf{RQ2}] How does compiler optimization affect LOHRE's recovery?
\item[\textbf{RQ3}] What is the practical quality of LOHRE-generated documentation, as judged by reverse-engineering practitioners?
\item[\textbf{RQ4}] What are the individual contributions of dynamic validation, EDH, and the knowledge base?
\item[\textbf{RQ5}] How does LOHRE's complete workflow compare with single-agent frameworks given the same tools?
\end{description}

\subsection{Experimental Setup}
\label{sec:setup}

\textbf{Datasets.} We use two complementary corpora.

(1)~\textbf{RealType}~\cite{dramko2025idioms}, a benchmark with UDT ground truth, well suited to general UDT recovery.
To ensure fairness, we match Idioms' setup exactly (inlining disabled).
RealType was compiled at four optimization levels (O0--O3, GNU Compiler Collection (GCC) 9.4.0).

(2)~\textbf{Windows RPC}, a dataset we assembled containing 4,280 exported and interface functions from 33 Windows~11 (build 26100.6899) System32 binaries.
We selected components using Microsoft Open Specifications~\cite{microsoftopenspecsrpc} to cover security, printing, networking, storage, and system management.
These components use handles and nested data structures shared across interface functions.
Recovering their complete type definitions therefore requires combining evidence from functions with separate roles, making the corpus suitable for studying scattered evidence.

To construct reference annotations, we first obtained version-matched program database (PDB) files from the Microsoft Symbol Server~\cite{microsoftsymbolserver}.
Their debug symbols identified functions and provided type information where available.
Because public PDBs may miss parameter and member types~\cite{microsoftpublicprivatesymbols}, we also consulted declarations in the Win32 API reference~\cite{microsoftwin32api} and public protocol definitions.
We checked these declarations against parameter use and member accesses in the disassembly, drawing on published technical analyses to clarify internal structures.

Public RPC specifications declare interface methods and their parameter types in Interface Definition Language (IDL).
The Microsoft IDL (MIDL) compiler generates client and server stub code to handle remote calls~\cite{microsoftmidl}.
Network Data Representation (NDR) metadata describes how this code encodes RPC arguments for transmission, providing evidence about parameter and aggregate representations~\cite{microsoftrpcndr}.

\textbf{Baselines.}
We compare LOHRE with three LLM-based methods that have runnable artifacts for our corpus.
Idioms~\cite{dramko2025idioms} is a fine-tuned neural decompiler that jointly predicts code and types.
For ReSym~\cite{RN22}, we use the aligned reproduction supplied by TRex~\cite{trex}.
The TRex authors independently verified the incompleteness of ReSym's original artifact.
LLM4Decompile~\cite{tan2024llm4decompile} generates decompiled code but does not explicitly emit structure definitions, so its UDT metrics are marked N/A.
We discuss other hybrid type-recovery systems in Section~\ref{sec:related_typerec}.
These comparisons evaluate complete systems, whose model backbones and inference budgets differ.

\textbf{Metrics.} We adopt Idioms' metrics~\cite{dramko2025idioms} to evaluate variable names, variable types, and UDT definitions.
\emph{Variable Name Accuracy} and \emph{Variable Type Accuracy} measure agreement with the reference variable names and types.
For UDTs, the scorer first aligns members by offset and width, then compares the structure size and recursive member types.
\emph{UDT Structural Accuracy} measures whether these layout requirements are met.
\emph{UDT Nominal Accuracy} additionally requires exact member-name matches.

All name comparisons use exact strings, including capitalization, underscores, and prefixes.
Consequently, a semantically equivalent name such as \texttt{readInput} does not match \texttt{read\_input}.
We retain this rule for consistent comparison with Idioms and ReSym, and use the user study (RQ3) to assess the documentation's semantic utility.

\textbf{Repeated runs.} Every accuracy value reported below is the mean of 8 independent runs.
These repetitions characterize run-to-run variability.

\textbf{Implementation and environment.}
LOHRE is implemented in Python~3.13.
In our experimental implementation, the Static Analysis MCP uses Ghidra~11.4.2~\cite{ghidra}, and the Dynamic Slice Emulation MCP uses Qiling~1.4.7~\cite{qiling}.
Both tools are accessed through FastMCP.
The agents use DeepSeek-V4-Flash~\cite{deepseekai2026deepseekv4}, and the knowledge base stores text entries separately for each binary.
All experiments ran on Ubuntu~22.04 with an Intel i9-10980XE processor and 128~GB random-access memory (RAM).
Two NVIDIA RTX 5060 GPUs were reserved for baseline inference.

\textbf{Execution budget and failure handling.}
Within the 20-iteration budget for each function, every baseline or mutated execution is limited to 100,000 instructions or five seconds.
We count an emulation as successful when execution reaches the designated return address, indicating that the function has returned to its caller.
Because a normal return may still report an application-level error, we record the return code separately.
We also distinguish exceptions, budget exhaustion, and initialization failures when recording unsuccessful executions.

When execution provides no usable dynamic evidence, the Executor retains the static hypotheses for documentation without treating environment failures as support for a type.
Any revised inputs or field tests remain within the same function's iteration budget.
For transient model-request failures, LOHRE makes at most three transmission attempts per request.

\begin{figure*}[!t]
    \centering
    \includegraphics[width=0.9\textwidth]{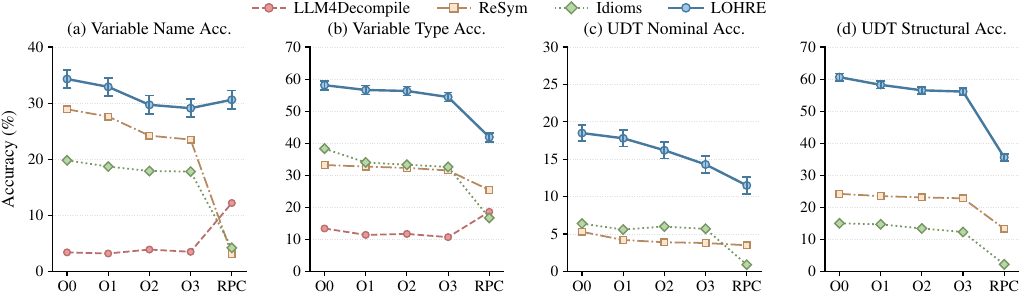}
    \caption{Recovery accuracy by dataset, optimization level, and metric.}
    \label{fig:overall_average}
\end{figure*}

\begin{table*}[t]
    \centering
    \caption{Type and name recovery accuracy on RealType and Windows RPC.}
    \label{tab:main_results}
    \resizebox{\textwidth}{!}{%
        \begin{tabular}{l @{\extracolsep{4pt}}cccc @{\extracolsep{4pt}}cccc @{\extracolsep{4pt}}cccc @{\extracolsep{4pt}}cccc}
            \toprule
            & \multicolumn{4}{c}{\textbf{Variable Name Acc. (\%)}} & \multicolumn{4}{c}{\textbf{Variable Type Acc. (\%)}} & \multicolumn{4}{c}{\textbf{UDT Nominal Acc. (\%)}} & \multicolumn{4}{c}{\textbf{UDT Structural Acc. (\%)}} \\
            \cmidrule{2-5} \cmidrule{6-9} \cmidrule{10-13} \cmidrule{14-17}
            \textbf{Dataset} & \textbf{LLM4Dec.} & \textbf{ReSym} & \textbf{Idioms} & \textbf{LOHRE} & \textbf{LLM4Dec.} & \textbf{ReSym} & \textbf{Idioms} & \textbf{LOHRE} & \textbf{LLM4Dec.} & \textbf{ReSym} & \textbf{Idioms} & \textbf{LOHRE} & \textbf{LLM4Dec.} & \textbf{ReSym} & \textbf{Idioms} & \textbf{LOHRE} \\
            \midrule
            \multicolumn{17}{c}{\textbf{Panel A: RealType}} \\
            \midrule
            RealType O0 & 3.4 & 28.9 & 19.8 & \textbf{34.3} & 13.4 & 33.2 & 38.3 & \textbf{58.1} & N/A & 5.3 & 6.4 & \textbf{18.5} & N/A & 24.2 & 15.0 & \textbf{60.6} \\
            RealType O1 & 3.2 & 27.6 & 18.7 & \textbf{32.9} & 11.4 & 32.7 & 34.0 & \textbf{56.6} & N/A & 4.2 & 5.6 & \textbf{17.8} & N/A & 23.5 & 14.7 & \textbf{58.2} \\
            RealType O2 & 3.9 & 24.2 & 17.9 & \textbf{29.7} & 11.7 & 32.3 & 33.3 & \textbf{56.3} & N/A & 3.9 & 6.0 & \textbf{16.2} & N/A & 23.1 & 13.4 & \textbf{56.5} \\
            RealType O3 & 3.5 & 23.5 & 17.8 & \textbf{29.1} & 10.7 & 31.5 & 32.6 & \textbf{54.4} & N/A & 3.8 & 5.7 & \textbf{14.3} & N/A & 22.8 & 12.3 & \textbf{56.1} \\
            \midrule
            \textbf{RealType Avg.} & 3.50 & 26.05 & 18.55 & \textbf{31.50} & 11.80 & 32.43 & 34.55 & \textbf{56.35} & N/A & 4.30 & 5.93 & \textbf{16.70} & N/A & 23.40 & 13.85 & \textbf{57.85} \\
            \textbf{Ratio} & 9.00 & 1.21 & 1.70 & -- & 4.78 & 1.74 & 1.63 & -- & N/A & 3.88 & 2.82 & -- & N/A & 2.47 & 4.18 & -- \\
            \midrule
            \multicolumn{17}{c}{\textbf{Panel B: Windows RPC}} \\
            \midrule
            Windows RPC & 12.2 & 3.1 & 4.2 & \textbf{30.6} & 18.6 & 25.4 & 16.7 & \textbf{41.9} & N/A & 3.5 & 0.9 & \textbf{11.5} & N/A & 13.3 & 2.2 & \textbf{35.6} \\
            \textbf{Ratio} & 2.51 & 9.87 & 7.29 & -- & 2.25 & 1.65 & 2.51 & -- & N/A & 3.29 & 12.78 & -- & N/A & 2.68 & 16.18 & -- \\
            \bottomrule
        \end{tabular}%
    }
\end{table*}

\subsection{Overall Performance (RQ1, RQ2)}
\label{sec:rq1}

Table~\ref{tab:main_results} reports the recovery results for each dataset.
Bold values mark the highest accuracy, and Ratio gives LOHRE's accuracy divided by that of the corresponding baseline.
N/A marks entries for which the comparison is not applicable.
Figure~\ref{fig:overall_average} plots these results by metric, with error bars showing one standard deviation above and below LOHRE's mean across eight runs.

\textbf{Run-to-run variability.}
Across the 20 combinations of evaluation setting and metric, the standard deviations over eight independent runs fall within the following ranges: $\sigma_{\text{UDT-Struct}} \in [0.7, 2.1]$ percentage points (pp), $\sigma_{\text{UDT-Nom}} \in [0.5, 1.8]$~pp, $\sigma_{\text{Var-Type}} \in [1.0, 2.9]$~pp, and $\sigma_{\text{Var-Name}} \in [1.1, 3.4]$~pp. These ranges characterize LOHRE's run-to-run variability and are small relative to the result gaps in Table~\ref{tab:main_results}.

\textbf{UDT and variable recovery (RQ1).}
LOHRE attains the highest accuracy in every reported evaluation setting and metric.
Across RealType O0--O3 and Windows RPC, LOHRE's mean Variable Name accuracy is 2.00$\times$ the mean for Idioms, 1.46$\times$ that for ReSym, and 5.98$\times$ that for LLM4Decompile.
For Variable Type accuracy, LOHRE achieves 1.73$\times$ the mean for Idioms, 1.72$\times$ the mean for ReSym, and 4.06$\times$ the mean for LLM4Decompile.
For UDT Nominal accuracy, LOHRE achieves 3.18$\times$ the mean for Idioms and 3.78$\times$ the mean for ReSym. For UDT Structural accuracy, it achieves 4.64$\times$ the mean for Idioms and 2.50$\times$ the mean for ReSym.
We attribute these gains mainly to iterative validation: execution provides evidence for identifying incorrect layouts and refining hypotheses when static code alone leaves them ambiguous.

\textbf{Robustness to compiler optimization (RQ2).}
On RealType, UDT Structural accuracy decreases from O0 to O3 for Idioms, ReSym, and LOHRE.
Idioms drops by 2.7~pp (18.0\% relative), ReSym by 1.4~pp (5.8\%), and LOHRE by 4.5~pp (7.4\%).
LOHRE's relative decrease lies between those of the two static baselines.
At O3, however, its accuracy of 56.1\% remains more than 2.4$\times$ that of the strongest baseline, ReSym at 22.8\%.
Thus, LOHRE retains its accuracy advantage under aggressive compiler optimization.

\textbf{Performance on Windows RPC.}
The Windows RPC dataset tests recovery when evidence about a UDT is distributed across functions.
LOHRE pairs a general-purpose LLM with MCP tools that test candidate layouts against execution behavior.
This lets prior knowledge guide candidate generation while execution provides evidence for refinement.
The training-free framework avoids the costs of collecting task-specific training data and adapting a model to the target domain.
Idioms and ReSym, whose training corpora omit Windows binaries, perform less well in this setting.
LOHRE attains accuracies of 30.6\%, 41.9\%, 11.5\%, and 35.6\% for Variable Name, Variable Type, UDT Nominal, and UDT Structural recovery, respectively, corresponding to 1.65$\times$--16.18$\times$ the accuracy of the evaluated baselines.
Recovery is harder than on RealType because larger UDTs contain rarely accessed fields, and UDT Nominal accuracy requires both a correct structure and exact field names.
A single mismatch can therefore cause a recovered definition to fail this metric.
Windows also revises internal structures across releases, so a candidate based on pretrained API knowledge may refer to an outdated layout.

Two results help explain how LOHRE's design contributes beyond pretrained API knowledge.
On RealType, which provides a GCC-based setting across four optimization levels, LOHRE reaches 57.85\% UDT Structural accuracy compared with Idioms' 13.85\%.
On Windows RPC, holding the LLM fixed while disabling the knowledge base reduces UDT Structural accuracy from 35.6\% to 10.0\% (Table~\ref{tab:ablation_results}).
These observations support the contribution of LOHRE's analysis workflow and cross-function evidence to recovery.

\textbf{Cost and efficiency.}
LOHRE incurs higher inference costs than fine-tuned baselines but requires no fine-tuning.
Table~\ref{tab:cost} summarizes function-level costs using the mean, standard deviation, 95th percentile (p95), and maximum of per-function token use and runtime.
It also includes LOHRE-NoEDH, the ablation of EDH's trace compression described in Section~\ref{sec:edh_ablation}.
For baseline token use, the Max column lists published context limits: 4096 for LLM4Decompile~\cite{tan2024llm4decompile} and Idioms~\cite{dramko2025idioms}, and 8192 for ReSym~\cite{RN22}.
These entries are capacity specifications, not measured per-function maxima.
At DeepSeek-V4-Flash list pricing, LOHRE's measured token use corresponds to an average cost of \$0.03 per function, with documentation produced in 401~s on average.
The p95 runtime and estimated cost are 879~s and \$0.05 per function, respectively.
These costs can be weighed against the effort required for manual analysis when assessing LOHRE's practical value.

\begin{table}[!t]
    \centering
    \caption{Per-function token use and runtime distributions.}
    \label{tab:cost}
    \resizebox{\columnwidth}{!}{%
    \begin{tabular}{l rrrr rrrr}
    \toprule
    & \multicolumn{4}{c}{\textbf{Tokens}} & \multicolumn{4}{c}{\textbf{Runtime (s)}} \\
    \cmidrule(lr){2-5}\cmidrule(lr){6-9}
    \textbf{System} & mean & sd & p95 & max & mean & sd & p95 & max \\
    \midrule
    LLM4Decompile  & 920 & 686 & 2376 & 4096 & 13.5 & 3.6 & 18.3 & 40.5 \\
    ReSym          & 2195 & 1295 & 4796 & 8192 & 2.6 & 1.5 & 5.8 & 11.3 \\
    Idioms         & 1500 & 783 & 3183 & 4096 & 1.9 & 1.0 & 3.7 & 8.7 \\
    \textbf{LOHRE} & 142k & 102k & 307k & 1179k & 401 & 266 & 879 & 2139 \\
    LOHRE-NoEDH    & 521k & 412k & 1274k & 3265k & 1183 & 814 & 2840 & 5763 \\
    \bottomrule
    \end{tabular}%
    }
\end{table}

\subsection{Documentation Quality (RQ3)}
\label{sec:user_study}

\begin{figure}[!t]
    \centering
    \includegraphics[width=0.95\columnwidth]{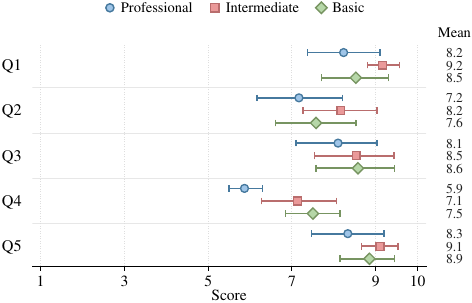}
    \caption{Documentation ratings across three participant groups.}
    \label{fig:user_study_results}
\end{figure}

We randomly selected ten LOHRE outputs from the evaluated set (five RealType and five Windows RPC) and ran a blind user study with ten participants of varying expertise and experience: three senior reverse engineers with more than three years of experience, three mid-level engineers with approximately two years of experience, and four practitioners with foundational reverse-engineering (RE) knowledge.
Following DeGPT~\cite{hu2024degpt}, each participant rated five aspects of each report on a 1--10 scale (1=completely wrong, 5=neutral, 10=accurate and complete).
This produced 500 aspect ratings (10$\times$10$\times$5).
Figure~\ref{fig:user_study_results} displays these aspects as Q1--Q5, corresponding to the function summary, function prototype, parameter descriptions, return-value semantics, and recovered UDT layouts, respectively.
Participation in the user study was voluntary, and all participants were informed that their anonymized responses would be used solely for this research.

Figure~\ref{fig:user_study_results} reports the mean ratings for each participant group, with horizontal lines indicating 95\% bootstrap confidence intervals.
Except for return-value analysis (Q4), all aspects score between 7.2 and 9.2.
The lower Q4 ratings reflect a coverage limitation: LOHRE currently infers return-value semantics from a single execution path and may miss alternative return points (Section~\ref{sec:limitations}).
Fleiss' $\kappa$ ranges from 0.56 to 0.65 across the five questions, indicating moderate to substantial inter-rater agreement.
Annotation guidelines and the full per-question rubrics are provided in the artifact.

\begin{table*}[!t]
    \centering
    \caption{Recovery accuracy with dynamic validation, the knowledge base, or EDH disabled.}
    \label{tab:ablation_results}
    \resizebox{\textwidth}{!}{%
        \begin{tabular}{l @{\extracolsep{4pt}}cccc @{\extracolsep{4pt}}cccc @{\extracolsep{4pt}}cccc @{\extracolsep{4pt}}cccc}
            \toprule
            & \multicolumn{4}{c}{\textbf{Variable Name Acc. (\%)}} & \multicolumn{4}{c}{\textbf{Variable Type Acc. (\%)}} & \multicolumn{4}{c}{\textbf{UDT Nominal Acc. (\%)}} & \multicolumn{4}{c}{\textbf{UDT Structural Acc. (\%)}} \\
            \cmidrule{2-5} \cmidrule{6-9} \cmidrule{10-13} \cmidrule{14-17}
            \textbf{Dataset} & \textbf{Static} & \textbf{NoKB} & \textbf{NoEDH} & \textbf{LOHRE} & \textbf{Static} & \textbf{NoKB} & \textbf{NoEDH} & \textbf{LOHRE} & \textbf{Static} & \textbf{NoKB} & \textbf{NoEDH} & \textbf{LOHRE} & \textbf{Static} & \textbf{NoKB} & \textbf{NoEDH} & \textbf{LOHRE} \\
            \midrule
            \multicolumn{17}{c}{\textbf{Panel A: RealType}} \\
            \midrule
            RealType O0 & 27.6 & 31.3 & 28.4 & \textbf{34.3} & 34.8 & 43.5 & 38.0 & \textbf{58.1} & 11.2 & 13.7 & 13.0 & \textbf{18.5} & 22.1 & 41.8 & 32.9 & \textbf{60.6} \\
            RealType O1 & 26.1 & 29.5 & 27.7 & \textbf{32.9} & 30.4 & 42.6 & 35.3 & \textbf{56.6} & 10.5 & 12.8 & 12.0 & \textbf{17.8} & 21.4 & 41.2 & 30.4 & \textbf{58.2} \\
            RealType O2 & 26.3 & 28.9 & 27.1 & \textbf{29.7} & 28.9 & 41.8 & 35.6 & \textbf{56.3} & 10.1 & 10.2 & 10.9 & \textbf{16.2} & 17.9 & 40.8 & 31.2 & \textbf{56.5} \\
            RealType O3 & 24.5 & 25.7 & 26.3 & \textbf{29.1} & 27.7 & 38.6 & 34.3 & \textbf{54.4} & 8.3 & 9.8 & 9.4 & \textbf{14.3} & 15.7 & 40.3 & 26.8 & \textbf{56.1} \\
            \midrule
            \textbf{RealType Avg.} & 26.13 & 28.85 & 27.38 & \textbf{31.50} & 30.45 & 41.63 & 35.80 & \textbf{56.35} & 10.03 & 11.63 & 11.33 & \textbf{16.70} & 19.28 & 41.03 & 30.33 & \textbf{57.85} \\
            \textbf{Ratio} & 1.21 & 1.09 & 1.15 & -- & 1.85 & 1.35 & 1.57 & -- & 1.67 & 1.44 & 1.47 & -- & 3.00 & 1.41 & 1.91 & -- \\
            \midrule
            \multicolumn{17}{c}{\textbf{Panel B: Windows RPC}} \\
            \midrule
            Windows RPC & 12.4 & 15.8 & 13.0 & \textbf{30.6} & 22.5 & 30.0 & 27.0 & \textbf{41.9} & 4.8 & 2.2 & 6.7 & \textbf{11.5} & 9.2 & 10.0 & 14.3 & \textbf{35.6} \\
            \textbf{Ratio} & 2.47 & 1.94 & 2.35 & -- & 1.86 & 1.40 & 1.55 & -- & 2.40 & 5.23 & 1.72 & -- & 3.87 & 3.56 & 2.49 & -- \\
            \bottomrule
        \end{tabular}%
    }
\end{table*}

\subsection{Ablation Study (RQ4)}
\label{sec:ablation}

\begin{figure}[!t]
    \centering
    \includegraphics[width=0.95\columnwidth]{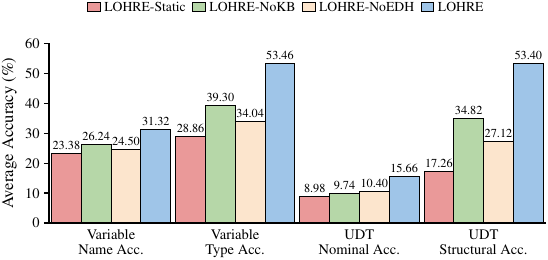}
    \caption{Mean ablation accuracy across the five evaluation settings.}
    \label{fig:ablation_fig}
\end{figure}

We construct three LOHRE variants to assess dynamic validation, cross-function knowledge reuse, and EDH compression.
(i)~\textbf{LOHRE-Static} disables dynamic validation entirely.
(ii)~\textbf{LOHRE-NoKB} disables the cross-function knowledge base.
(iii)~\textbf{LOHRE-NoEDH} retains dynamic validation but replaces EDH's compressed feedback with full execution traces supplied to the Executor.
In LOHRE-NoEDH, the multi-agent framework, knowledge base, and mutation strategy remain the same.
All variants use the same evaluation settings as full LOHRE: RealType at O0--O3 and Windows RPC, with eight runs per setting.
Table~\ref{tab:ablation_results} reports mean accuracy over eight runs for each evaluation setting.
Figure~\ref{fig:ablation_fig} averages those means across five settings: RealType at O0--O3 and Windows RPC.

\textbf{Dynamic validation.}
Disabling dynamic validation (LOHRE-Static) reduces UDT Structural accuracy by 66.7\% on RealType (57.85\%$\rightarrow$19.28\%) and 74.2\% on Windows RPC (35.6\%$\rightarrow$9.2\%).
The larger relative drop on Windows RPC is consistent with the interaction between validation and knowledge reuse.
Execution feedback both helps identify incorrect hypotheses and provides evidence for combining UDT information across functions.
Without it, errors in static hypotheses can persist during aggregation.

\textbf{Knowledge base.}
Disabling the knowledge base (LOHRE-NoKB) reduces UDT Structural accuracy from 57.85\% to 41.03\% on RealType, a 29\% drop.
The loss is greater on Windows RPC, where accuracy falls from 35.6\% to 10.0\%, a 72\% drop.
RealType UDTs are largely local to individual functions, whereas Windows RPC distributes information about shared UDTs across functions.
This contrast supports the role of the knowledge base in combining scattered evidence.

\textbf{Execution-Depth Heuristic.}
\label{sec:edh_ablation}
Replacing EDH's compressed feedback with full traces (LOHRE-NoEDH) reduces recovery accuracy on both datasets.
On RealType, UDT Structural, UDT Nominal, Variable Type, and Variable Name accuracies are 30.33\%, 11.33\%, 35.80\%, and 27.38\%, respectively.
LOHRE-Static obtains 19.28\% UDT Structural, 10.03\% UDT Nominal, 30.45\% Variable Type, and 26.13\% Variable Name accuracy; full LOHRE obtains 57.85\%, 16.70\%, 56.35\%, and 31.50\% on the same metrics, respectively.
On Windows RPC, LOHRE-NoEDH obtains 14.3\% UDT Structural, 6.7\% UDT Nominal, 27.0\% Variable Type, and 13.0\% Variable Name accuracy; full LOHRE obtains 35.6\%, 11.5\%, 41.9\%, and 30.6\% on the same metrics, respectively.
Table~\ref{tab:cost} also shows a higher cost: mean token use reaches 521k and mean runtime reaches 1183~s, corresponding to 3.7$\times$ the token use and 3.0$\times$ the runtime of full LOHRE.
We attribute this accuracy loss to the difficulty of reasoning over long execution histories~\cite{gao2026ink,zeng2026loca}.
This explanation is supported by studies showing that long inputs can impair the use of relevant evidence even within a model's supported context window~\cite{liu2024lost,modarressi2025nolima,du2025context}.
At the same time, carrying full traces into subsequent turns increases token costs by repeatedly including the accumulated execution history~\cite{kim2026defensive}.
EDH retains the depth comparison and the field behavior needed to interpret it in a compact validation interface.
This compression keeps the evidence for the selected hypothesis within a concise working context.

\subsection{LOHRE vs. Single-Agent Framework (RQ5)}
\label{sec:rq5}

We compare LOHRE with two single-agent baselines on RealType-O3, both equipped with the same Static Analysis MCP and Dynamic Slice Emulation MCP tools as LOHRE. 
The \textbf{DeepSeek Harness}~\cite{deepseekharness} is an execution harness specifically designed for DeepSeek models, using native tool invocation on the same DeepSeek-V4-Flash~\cite{deepseekai2026deepseekv4} backbone as LOHRE but without LOHRE's hypothesize-validate-refine loop. 
\textbf{Claude Code} is a representative state-of-the-art general-purpose agent framework running in its default configuration on the same task and toolset but with a \emph{different} backbone (Claude Sonnet 4.6~\cite{anthropic2026claudesonnet46}).
Table~\ref{tab:framework_vs_model} reports the results, with parenthesized values giving LOHRE's accuracy divided by that of the corresponding baseline.

\begin{table}[!t]
\centering
\caption{Recovery accuracy of LOHRE and general-purpose agents on RealType-O3.}
\label{tab:framework_vs_model}
\resizebox{\columnwidth}{!}{%
\begin{tabular}{lcccc}
\toprule
\textbf{Configuration} & \textbf{Name} & \textbf{Type} & \textbf{UDT Nom.} & \textbf{UDT Struct.} \\
\midrule
DeepSeek Harness     & 24.9 (1.17$\times$) & 33.2 (1.64$\times$) & 7.90 (1.81$\times$) & 21.6 (2.60$\times$) \\
Claude Code        & 27.2 (1.07$\times$) & 48.5 (1.12$\times$) & 10.3 (1.39$\times$) & 40.1 (1.40$\times$) \\
\midrule
\textbf{LOHRE}     & \textbf{29.1}       & \textbf{54.4}       & \textbf{14.3}       & \textbf{56.1}       \\
\bottomrule
\end{tabular}%
}
\end{table}

LOHRE achieves higher accuracy than both single-agent baselines on every metric while using fewer tokens.
On RealType-O3, the DeepSeek Harness uses the same DeepSeek-V4-Flash backbone as LOHRE and consumes 633k tokens per function on average, compared with LOHRE's 126k, meaning that it uses 5.01$\times$ as many tokens.
This comparison evaluates LOHRE's complete workflow and agent organization with the model backbone held fixed.
Claude Code provides a complementary comparison with a general-purpose agent framework using Claude Sonnet 4.6.
It consumes 571k tokens per function (4.52$\times$ LOHRE's use) and has lower accuracy on every metric, including UDT Structural accuracy of 40.1 versus LOHRE's 56.1.
Together, these results support the value of task-specific hypothesize-validate-refine orchestration.

Context management provides an explanation for the token difference.
During dynamic analysis, detailed traces and operational history accumulate, and repeatedly including them in later requests increases token costs~\cite{kim2026defensive}.
Long inputs can also impair the use of relevant evidence~\cite{liu2024lost,modarressi2025nolima,du2025context}.
LOHRE supplies each agent with context focused on its task, limiting the history carried into subsequent requests.
This task-focused context may help explain why the same-backbone DeepSeek Harness uses 5.01$\times$ as many tokens as LOHRE.

\section{Discussion}
\label{sec:discussion}

\subsection{Why the Paradigm Works}
\label{sec:why}

LOHRE is built around a central shift: it treats LLM outputs as falsifiable hypotheses rather than authoritative predictions, using execution feedback to identify inconsistencies and provide refinement signals rather than to prove type correctness.
LOHRE tests competing interpretations of the selected field using paired inputs, and EDH compresses the traces into a depth comparison and a summary of the field's observed uses.
The Executor uses this feedback to evaluate the candidate's plausibility and carry the relevant evidence into subsequent iterations.
The lower accuracy of the static-only ablation supports the contribution of dynamic validation to recovery.

The knowledge base extends this process across functions that use the same UDT.
It carries field hypotheses and their confidences forward, so later tests can concentrate on uncertain fields and extend the recovered definition.
If detailed analysis links an execution anomaly to an inherited field, the Executor lowers that field's confidence and reexamines its interpretation.
Dynamic validation and knowledge reuse thus work together: execution supplies empirical evidence for refinement, while the knowledge base lets later functions build on and revise earlier hypotheses.
This combination is particularly useful in Windows RPC, where functions without direct call relationships often access different parts of a shared UDT.

\subsection{Coverage Boundary and Heuristic Robustness}
\label{sec:limitations}

Dynamic validation requires a suitable execution environment for the target function.
On RealType, emulation succeeds in 83.8\% of cases.
In the remaining 16.2\%, emulation fails because of runtime-environment problems, such as missing \texttt{.so} dependencies, even though their source code compiles successfully.
These cases use static-only analysis and remain in the evaluation, so their results contribute to the reported accuracy.
Programs whose source code cannot be compiled are excluded before evaluation.

\begin{figure}[!t]
    \centering
    \includegraphics[width=0.95\columnwidth]{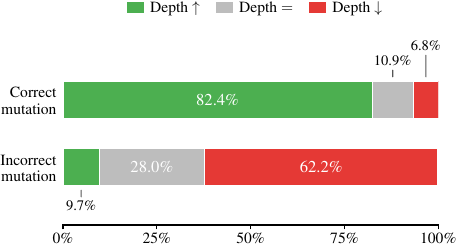}
    \caption{Distribution of paired-run depth changes on RealType.}
    \label{fig:heuristic}
\end{figure}

Even when execution is possible, a greater depth can cause EDH to favor an incorrect input if that input enters an error path covering many blocks.
Counting each block once within the target module limits the influence of repeated execution and external library internals, but application-level error paths can still reach more blocks.
EDH's behavior summary helps the Executor interpret this depth preference.
Figure~\ref{fig:heuristic} shows that correct mutations increase depth in 82.4\% of RealType cases, while incorrect mutations decrease it in 62.2\%.
The 9.7\% misdirection rate arises when incorrect mutations enter deeper application-level error paths.
LOHRE reexamines and corrects the affected hypotheses in 1.61 refinement iterations on average.
For the 28.0\% of incorrect mutations with unchanged depth, the Executor uses the behavior summary and relevant static code to guide further refinement.

When a field affects data without changing control flow, its observed uses can still constrain its type, although they may not establish the precise source declaration~\cite{trex}.
For unaccessed fields, neighboring accesses may reveal an interior gap without identifying its constituent members.
Unaccessed trailing fields may leave no evidence of their extent.

Coverage also limits parameter and return-value recovery: unused parameters may be omitted from recovered signatures, and the single-path return-value analysis in Section~\ref{sec:user_study} can miss alternative return points.
Even when the relevant behavior is observed, recovered names may describe a field's role while differing from its source identifier, affecting UDT Nominal accuracy under exact-match scoring.
The artifact presents these recovery cases.

\subsection{Threats to Validity}
\label{sec:threats}

The evaluation measures exact agreement with reference names and types, so semantically equivalent names can still be scored as incorrect.
The user study complements these metrics with practitioners' judgments of the documentation, and Fleiss' $\kappa$ characterizes agreement among their ratings.

The eight repeated runs use a fixed DeepSeek-V4-Flash model.
The observed standard-deviation bound ($\leq 3.4$~pp) therefore describes run-to-run variation only under this configuration.
A different model checkpoint or decoding configuration may change variability.

The datasets cover complementary settings but impose limits on generalization.
RealType excludes Microsoft APIs, although overlap with model training data remains possible.
Its four optimization levels share one source corpus and compiler, so they test optimization robustness rather than corpus diversity.
Windows RPC draws on Microsoft Open Specifications to cover a representative range of protocol components.
Dynamic validation is available for the 83.8\% of RealType cases in which emulation succeeds.
The remaining 16.2\% use static-only analysis and remain in the evaluation.

\section{Related Work}
\label{sec:related}
\label{sec:related_neural}
\label{sec:related_typerec}
\label{sec:related_llm}

LLMs support program understanding and security analysis~\cite{jelodar2025llmsecurity,lea2025rex86,sha2025llasm,hussain2025vulbinllm,11334428,liu2025llmtaint}, with approaches ranging from interactive querying~\cite{pearce2022popquiz} to agent-based orchestration~\cite{chen2025recopilot,shao2025craken,mahaloz2024daila}.
Automated testing complements this analysis by generating and selecting inputs that expose program behavior~\cite{RN21,zhang2026thompson,cui2026visiongui}.
LOHRE connects analysis orchestration with test generation in a hypothesize-validate-refine loop: paired runs challenge UDT hypotheses, and EDH summarizes feedback for refinement.

For code reconstruction, neural decompilation recovers high-level source code from binaries~\cite{tan2024llm4decompile,jiang2023nova}, with extensions for particular platforms and deobfuscation~\cite{she2024wadec,liu2025armqwen2,choi2024chatdeob}.
Other methods improve decompiler output through identifier recovery, commenting, and restructuring~\cite{hu2024degpt,xu2023lmpa,11334626,lacomis2019dire,pal2024varbert,chen2022dirty}, or jointly recover code and data structures~\cite{dramko2025idioms,wong2025decllm,feng2025refdecompile}.
CLAP aligns assembly with natural-language explanations~\cite{wang2024clap}, while LLM Compiler learns assembly and compiler intermediate representations~\cite{cummins2025llmcompiler}.
Both show how useful semantics can be inferred from low-level code.
These advances make binaries easier to interpret, but plausible static predictions can still contain incorrect layouts~\cite{hu2025sok}.

Type-recovery methods complement code reconstruction through constraint-based inference~\cite{lee2011tie,noonan2016retypd,smith2025binsub}, probabilistic and neural techniques~\cite{zhang2021osprey,zhuo2025beyondc,299717,he2018debin,chen2022dirty,li2026bytetr}, and structure inference from observed executions~\cite{lin2010rewards,slowinska2011howard}.
For quantitative comparison, we evaluate LOHRE against ReSym~\cite{RN22}, Idioms~\cite{dramko2025idioms}, and LLM4Decompile~\cite{tan2024llm4decompile} because runnable implementations are available for our corpus.
TypeForge~\cite{wang2025typeforge} ranks constraint-generated types with an LLM but requires decompiler-inferred standard-library signatures unavailable in our setting.
Idioms also reports that TypeForge cannot run on the RealType-derived corpus~\cite{dramko2025idioms}.
HyRES~\cite{sha2025hyres} supplies LOHRE's aggregation heuristics, but its public artifact underspecifies normalization, constraint rules, and aggregation settings, preventing a controlled comparison.
In modular binaries, a shared UDT may be accessed by functions without direct call relationships, leaving each function with only part of the layout evidence.
LOHRE therefore aggregates evidence across these functions and uses execution to refine the resulting UDT hypotheses.
It carries field confidences forward to focus testing on uncertain fields and revisits inherited hypotheses when execution analysis reveals conflicts.

\section{Conclusion}
\label{sec:conclusion}

We presented LOHRE, an LLM-orchestrated hybrid analysis framework for execution-grounded type recovery.
In LOHRE, paired dynamic slice emulation tests candidate UDT layouts, EDH compresses the resulting execution feedback, and a cross-function knowledge base consolidates compatible evidence.
Across RealType and Windows RPC, LOHRE achieves UDT Structural accuracies of 57.85\% and 35.6\%, respectively, outperforming the evaluated static baselines by 2.47--4.18$\times$ and 2.68--16.18$\times$.
A practitioner study (Fleiss' $\kappa$ 0.56--0.65) indicates that the documentation is perceived as useful.
The EDH ablation further shows that compressed feedback supports higher recovery accuracy and lower token use and runtime than full execution traces.
Future work will extend EDH's field behavior summaries to cover external library implementations and evaluate portability across additional LLMs and architectures.
The hypothesize-validate-refine loop could also support function-prototype or protocol-state recovery when candidate interpretations can be tested through observable behavior.


\balance
\bibliographystyle{IEEEtran}
\bibliography{references}

@inproceedings{shoshitaishvili2016sok,
  author        = {Shoshitaishvili, Yan and Wang, Ruoyu and Salls, Christopher and Stephens, Nick and Polino, Mario and Dutcher, Andrew and Grosen, John and Feng, Siji and Hauser, Christophe and Kruegel, Christopher and Vigna, Giovanni},
  title         = {{SoK: (State of) The Art of War: Offensive Techniques in Binary Analysis}},
  booktitle     = {Proceedings of the 2016 IEEE Symposium on Security and Privacy},
  pages         = {138--157},
  year          = {2016}
}

@misc{hu2025sok,
  author        = {Hu, Xinyu and Fu, Zhiwei and Xie, Shaocong and Ding, Steven H. H. and Charland, Philippe},
  title         = {{SoK: Potentials and Challenges of Large Language Models for Reverse Engineering}},
  year          = {2025},
  howpublished  = {\url{https://arxiv.org/abs/2509.21821}},
  eprint        = {2509.21821}
}

@misc{ghidra,
  title         = {{Ghidra}},
  howpublished  = {\url{https://ghidra-sre.org}},
  note          = {Accessed in 2026}
}

@inproceedings{hu2024degpt,
  author        = {Hu, Peiwei and Liang, Ruigang and Chen, Kai},
  title         = {{DeGPT: Optimizing Decompiler Output with LLM}},
  booktitle     = {Proceedings of the 2024 Network and Distributed System Security Symposium},
  year          = {2024}
}

@inproceedings{tan2024llm4decompile,
  author        = {Tan, Hanzhuo and Luo, Qi and Li, Jing and Zhang, Yuqun},
  title         = {{LLM}4{D}ecompile: Decompiling Binary Code with Large Language Models},
  booktitle     = {Proceedings of the 2024 Conference on Empirical Methods in Natural Language Processing},
  pages         = {3473--3487},
  year          = {2024}
}

@misc{qiling,
  title         = {{Qiling Framework}},
  howpublished  = {\url{https://github.com/qilingframework/qiling}},
  note          = {Accessed in 2025}
}

@inproceedings{RN21,
  author        = {Srivastava, Prashast and Payer, Mathias},
  title         = {Gramatron: effective grammar-aware fuzzing},
  booktitle     = {Proceedings of the 30th ACM SIGSOFT International Symposium on Software Testing and Analysis},
  pages         = {244--256},
  year          = {2021}
}

@inproceedings{RN22,
  author        = {Xie, Danning and Zhang, Zhuo and Jiang, Nan and Xu, Xiangzhe and Tan, Lin and Zhang, Xiangyu},
  title         = {ReSym: Harnessing LLMs to Recover Variable and Data Structure Symbols from Stripped Binaries},
  booktitle     = {Proceedings of the 2024 ACM SIGSAC Conference on Computer and Communications Security},
  pages         = {4554--4568},
  year          = {2024}
}

@misc{hussain2025vulbinllm,
  author        = {Hussain, Nasir and Chen, Haohan and Tran, Chanh and Huang, Philip and Li, Zhuohao and Chugh, Pravir and Chen, William and Kundu, Ashish and Tian, Yuan},
  title         = {{VulBinLLM: LLM-powered Vulnerability Detection for Stripped Binaries}},
  year          = {2025},
  howpublished  = {\url{https://arxiv.org/abs/2505.22010}},
  eprint        = {2505.22010}
}

@misc{jelodar2025llmsecurity,
  author        = {Jelodar, Hamed and Bai, Samita and Hamedi, Parisa and Mohammadian, Hesamodin and Razavi-Far, Roozbeh and Ghorbani, Ali},
  title         = {{Large Language Model (LLM) for Software Security: Code Analysis, Malware Analysis, Reverse Engineering}},
  year          = {2025},
  howpublished  = {\url{https://arxiv.org/abs/2504.07137}},
  eprint        = {2504.07137}
}

@inproceedings{she2024wadec,
  author        = {She, Xinyu and Zhao, Yanjie and Wang, Haoyu},
  title         = {{WaDec}: Decompiling {WebAssembly} Using Large Language Model},
  booktitle     = {Proceedings of the 39th IEEE/ACM International Conference on Automated Software Engineering},
  pages         = {481--492},
  year          = {2024}
}

@inproceedings{wang2025typeforge,
  author        = {Wang, Yanzhong and Liang, Ruigang and Li, Yilin and Hu, Peiwei and Chen, Kai and Zhang, Bolun},
  title         = {{TypeForge: Synthesizing and Selecting Best-Fit Composite Data Types for Stripped Binaries}},
  booktitle     = {Proceedings of the 2025 IEEE Symposium on Security and Privacy},
  pages         = {2847--2864},
  year          = {2025}
}

@inproceedings{xu2023lmpa,
  author        = {Xu, Xiangzhe and Zhang, Zhuo and Su, Zian and Huang, Ziyang and Feng, Shiwei and Ye, Yapeng and Jiang, Nan and Xie, Danning and Cheng, Siyuan and Tan, Lin and Zhang, Xiangyu},
  title         = {{Unleashing the Power of Generative Model in Recovering Variable Names from Stripped Binary}},
  booktitle     = {Proceedings of the 32nd Network and Distributed System Security Symposium},
  year          = {2025}
}

@inproceedings{jiang2023nova,
  author        = {Jiang, Nan and Wang, Chengxiao and Liu, Kevin and Xu, Xiangzhe and Tan, Lin and Zhang, Xiangyu and Babkin, Petr},
  title         = {{Nova: Generative Language Models for Assembly Code with Hierarchical Attention and Contrastive Learning}},
  booktitle     = {Proceedings of the 13th International Conference on Learning Representations},
  year          = {2025}
}

@article{wong2025decllm,
  author        = {Wong, Wai Kin and Wu, Daoyuan and Wang, Huaijin and Li, Zongjie and Liu, Zhibo and Wang, Shuai and Tang, Qiyi and Nie, Sen and Wu, Shi},
  title         = {{DecLLM: LLM-Augmented Recompilable Decompilation for Enabling Programmatic Use of Decompiled Code}},
  journal       = {Proceedings of the ACM on Software Engineering},
  volume        = {2},
  pages         = {1841--1864},
  year          = {2025}
}

@misc{feng2025refdecompile,
  author        = {Feng, Yunlong and Li, Bohan and Shi, Xiaoming and Zhu, Qingfu and Che, Wanxiang},
  title         = {{ReF Decompile: Relabeling and Function Call Enhanced Decompile}},
  year          = {2025},
  howpublished  = {\url{https://arxiv.org/abs/2502.12221}},
  eprint        = {2502.12221}
}

@inproceedings{choi2024chatdeob,
  author        = {Choi, Byunggeon and Jin, Hongjoo and Lee, Dong Hoon and Choi, Wonsuk},
  title         = {{ChatDEOB: An Effective Deobfuscation Method Based on Large Language Model}},
  booktitle     = {Proceedings of the 25th International Conference on Information Security Applications},
  pages         = {151--163},
  year          = {2025}
}

@inproceedings{liu2025armqwen2,
  author        = {Liu, Jiahan and Lin, Jian and Liu, Haoran and Wang, Yonghao and Jing, Jing},
  title         = {{ARMQwen2: Enhancing C Language Decompilation on ARM Platform Using Large Language Model}},
  booktitle     = {Proceedings of the 21st International Conference on Intelligent Computing},
  pages         = {221--234},
  year          = {2025}
}

@article{sha2025hyres,
  author        = {Sha, Zihan and Shu, Hui and Wang, Hao and Gao, Zeyu and Lan, Yang and Zhang, Chao},
  title         = {{HyRES: Recovering Data Structures in Binaries via Semantic Enhanced Hybrid Reasoning}},
  journal       = {ACM Transactions on Software Engineering and Methodology},
  volume        = {35},
  pages         = {1--28},
  year          = {2026}
}

@inproceedings{zhang2021osprey,
  author        = {Zhang, Zhuo and Ye, Yapeng and You, Wei and Tao, Guanhong and Lee, Wen-chuan and Kwon, Yonghwi and Aafer, Yousra and Zhang, Xiangyu},
  title         = {Osprey: Recovery of variable and data structure via probabilistic analysis for stripped binary},
  booktitle     = {Proceedings of the 2021 IEEE Symposium on Security and Privacy},
  pages         = {813--832},
  year          = {2021}
}

@misc{zhuo2025beyondc,
  author        = {Zhuo, Zhuo and Zhang, Xiangyu},
  title         = {{Beyond C/C++}: Probabilistic and {LLM} Methods for Next-Generation Software Reverse Engineering},
  year          = {2025},
  howpublished  = {\url{https://arxiv.org/abs/2506.03504}},
  eprint        = {2506.03504}
}

@misc{pearce2022popquiz,
  author        = {Pearce, Hammond and Tan, Benjamin and Krishnamurthy, Prashanth and Khorrami, Farshad and Karri, Ramesh and Dolan-Gavitt, Brendan},
  title         = {{Pop Quiz! Can a Large Language Model Help with Reverse Engineering?}},
  year          = {2022},
  howpublished  = {\url{https://arxiv.org/abs/2202.01142}},
  eprint        = {2202.01142}
}

@article{liu2025llmtaint,
  author        = {Liu, Puzhuo and Sun, Chengnian and Zheng, Yaowen and Feng, Xuan and Qin, Chuan and Wang, Yuncheng and Xu, Zhenyang and Li, Zhi and Di, Peng and Jiang, Yu and Sun, Limin},
  title         = {{LLM-Powered Static Binary Taint Analysis}},
  journal       = {ACM Transactions on Software Engineering and Methodology},
  volume        = {34},
  pages         = {1--36},
  year          = {2025}
}

@misc{chen2025recopilot,
  author        = {Chen, Guoqiang and Sun, Huiqi and Liu, Daguang and Wang, Zhiqi and Wang, Qiang and Yin, Bin and Liu, Lu and Ying, Lingyun},
  title         = {{ReCopilot: Reverse Engineering Copilot in Binary Analysis}},
  year          = {2025},
  howpublished  = {\url{https://arxiv.org/abs/2505.16366}},
  eprint        = {2505.16366}
}

@article{sha2025llasm,
  author        = {Sha, Zihan and Wang, Hao and Gao, Zeyu and Shu, Hui and Zhang, Bolun and Wang, Ziqing and Zhang, Chao},
  title         = {{llasm: Naming functions in binaries by fusing encoder-only and decoder-only LLMs}},
  journal       = {ACM Transactions on Software Engineering and Methodology},
  volume        = {34},
  pages         = {1--22},
  year          = {2025}
}

@misc{shao2025craken,
  author        = {Shao, Minghao and Xi, Haoran and Rani, Nanda and Udeshi, Meet and Putrevu, Venkata Sai Charan and Milner, Kimberly and Dolan-Gavitt, Brendan and Shukla, Sandeep Kumar and Krishnamurthy, Prashanth and Khorrami, Farshad and Karri, Ramesh and Shafique, Muhammad},
  title         = {{CRAKEN}: Cybersecurity {LLM} agent with knowledge-based execution},
  year          = {2025},
  howpublished  = {\url{https://arxiv.org/abs/2505.17107}},
  eprint        = {2505.17107}
}

@misc{mahaloz2024daila,
  title         = {{DAILA: Decompiler Artificial Intelligence Language Assistant}},
  howpublished  = {\url{https://github.com/mahaloz/DAILA}},
  note          = {Accessed in 2026}
}

@misc{mcp,
  title         = {{Model Context Protocol Specification}},
  howpublished  = {\url{https://modelcontextprotocol.io/specification/2025-06-18}},
  note          = {Accessed in 2026}
}

@misc{li2023towards,
  author        = {Li, Zehan and Zhang, Xin and Zhang, Yanzhao and Long, Dingkun and Xie, Pengjun and Zhang, Meishan},
  title         = {Towards general text embeddings with multi-stage contrastive learning},
  year          = {2023},
  howpublished  = {\url{https://arxiv.org/abs/2308.03281}},
  eprint        = {2308.03281}
}

@inproceedings{299717,
  author        = {Chang Zhu and Ziyang Li and Anton Xue and Ati Priya Bajaj and Wil Gibbs and Yibo Liu and Rajeev Alur and Tiffany Bao and Hanjun Dai and Adam Doup{\'e} and Mayur Naik and Yan Shoshitaishvili and Ruoyu Wang and Aravind Machiry},
  title         = {{TYGR}: Type Inference on Stripped Binaries using Graph Neural Networks},
  booktitle     = {Proceedings of the 33rd USENIX Security Symposium},
  pages         = {4283--4300},
  year          = {2024}
}

@inproceedings{lea2025rex86,
  author        = {Lea, Darrin and Ghawaly, James and Richard III, Golden and Ali-Gombe, Aisha and Case, Andrew},
  title         = {{REx86: A Local Large Language Model for Assisting in x86 Assembly Reverse Engineering}},
  booktitle     = {Proceedings of the 41st IEEE Annual Computer Security Applications Conference},
  pages         = {108--122},
  year          = {2025}
}

@inproceedings{11334626,
  author        = {Li, Gangyang and Shang, Xiuwei and Cheng, Shaoyin and Zhang, Junqi and Hu, Li and Zhu, Xu and Zhang, Weiming and Yu, Nenghai},
  title         = {PseudoFix: Refactoring Distorted Structures in Decompiled C Pseudocode},
  booktitle     = {Proceedings of the 40th IEEE/ACM International Conference on Automated Software Engineering},
  pages         = {841--853},
  year          = {2025}
}

@inproceedings{11334428,
  author        = {Zhang, Yiran and Xu, Zhengzi and Lang, Zhe and Liu, Chengyue and Sun, Yuqiang and Guo, Wenbo and Liu, Chengwei and Sun, Weisong and Liu, Yang},
  title         = {BinStruct: Binary Structure Recovery Combining Static Analysis and Semantics},
  booktitle     = {Proceedings of the 40th IEEE/ACM International Conference on Automated Software Engineering},
  pages         = {648--660},
  year          = {2025}
}

@inproceedings{trex,
  author        = {Bosamiya, Jay and Woo, Maverick and Parno, Bryan},
  title         = {{TRex}: Practical Type Reconstruction for Binary Code},
  booktitle     = {Proceedings of the 34th USENIX Security Symposium},
  pages         = {6897--6915},
  year          = {2025}
}

@misc{kim2026defensive,
  author        = {Kim, Hyunjun},
  title         = {Defensive M2S: Training Guardrail Models on Compressed Multi-turn Conversations},
  year          = {2026},
  howpublished  = {\url{https://arxiv.org/abs/2601.00454}},
  eprint        = {2601.00454}
}

@inproceedings{dramko2025idioms,
  author        = {Dramko, Luke and Le Goues, Claire and Schwartz, Edward J.},
  title         = {Idioms: A Simple and Effective Framework for Turbo-Charging Local Neural Decompilation with Well-Defined Types},
  booktitle     = {Proceedings of the 33rd Network and Distributed System Security Symposium},
  year          = {2026}
}

@misc{deepseekai2026deepseekv4,
  author        = {{DeepSeek-AI}},
  title         = {{DeepSeek-V4: Towards Highly Efficient Million-Token Context Intelligence}},
  year          = {2026},
  howpublished  = {\url{https://arxiv.org/abs/2606.19348}},
  eprint        = {2606.19348}
}

@misc{anthropic2026claudesonnet46,
  title         = {{Claude Sonnet 4.6 System Card}},
  howpublished  = {\url{https://www.anthropic.com/claude-sonnet-4-6-system-card}},
  note          = {Accessed in 2026}
}

@inproceedings{lee2011tie,
  author        = {Lee, JongHyup and Avgerinos, Thanassis and Brumley, David},
  title         = {{TIE}: Principled Reverse Engineering of Types in Binary Programs},
  booktitle     = {Proceedings of the 18th Network and Distributed System Security Symposium},
  year          = {2011}
}

@inproceedings{noonan2016retypd,
  author        = {Noonan, Matt and Loginov, Alexey and Cok, David},
  title         = {Polymorphic Type Inference for Machine Code},
  booktitle     = {Proceedings of the 37th ACM SIGPLAN Conference on Programming Language Design and Implementation},
  pages         = {27--41},
  year          = {2016}
}

@inproceedings{smith2025binsub,
  author        = {Smith, Ian},
  title         = {{BinSub}: The Simple Essence of Polymorphic Type Inference for Machine Code},
  booktitle     = {Proceedings of the 31st International Static Analysis Symposium},
  pages         = {425--450},
  year          = {2025}
}

@inproceedings{he2018debin,
  author        = {He, Jingxuan and Ivanov, Pesho and Tsankov, Petar and Raychev, Veselin and Vechev, Martin},
  title         = {{Debin}: Predicting Debug Information in Stripped Binaries},
  booktitle     = {Proceedings of the 2018 ACM SIGSAC Conference on Computer and Communications Security},
  pages         = {1667--1680},
  year          = {2018}
}

@inproceedings{chen2022dirty,
  author        = {Chen, Qibin and Lacomis, Jeremy and Schwartz, Edward J. and Le Goues, Claire and Neubig, Graham and Vasilescu, Bogdan},
  title         = {Augmenting Decompiler Output with Learned Variable Names and Types},
  booktitle     = {Proceedings of the 31st USENIX Security Symposium},
  pages         = {4327--4343},
  year          = {2022}
}

@article{li2026bytetr,
  author        = {Li, Gangyang and Shang, Xiuwei and Cheng, Shaoyin and Zhang, Junqi and Hu, Li and Zhu, Xu and Zhang, Weiming and Yu, Nenghai},
  title         = {Beyond the Edge of Function: Unraveling the Patterns of Type Recovery in Binary Code},
  journal       = {ACM Transactions on Software Engineering and Methodology},
  year          = {2026}
}

@inproceedings{lacomis2019dire,
  author        = {Lacomis, Jeremy and Yin, Pengcheng and Schwartz, Edward J. and Allamanis, Miltiadis and Le Goues, Claire and Neubig, Graham and Vasilescu, Bogdan},
  title         = {{DIRE}: A Neural Approach to Decompiled Identifier Naming},
  booktitle     = {Proceedings of the 34th IEEE/ACM International Conference on Automated Software Engineering},
  pages         = {628--639},
  year          = {2019}
}

@inproceedings{pal2024varbert,
  author        = {Pal, Kuntal Kumar and Bajaj, Ati Priya and Banerjee, Pratyay and Dutcher, Audrey and Nakamura, Mutsumi and Basque, Zion Leonahenahe and Gupta, Himanshu and Sawant, Saurabh Arjun and Anantheswaran, Ujjwala and Shoshitaishvili, Yan and Doup{\'e}, Adam and Baral, Chitta and Wang, Ruoyu},
  title         = {`Len or Index or Count, Anything but v1': Predicting Variable Names in Decompilation Output with Transfer Learning},
  booktitle     = {Proceedings of the 2024 IEEE Symposium on Security and Privacy},
  pages         = {4069--4087},
  year          = {2024}
}

@inproceedings{lin2010rewards,
  author        = {Lin, Zhiqiang and Zhang, Xiangyu and Xu, Dongyan},
  title         = {Automatic Reverse Engineering of Data Structures from Binary Execution},
  booktitle     = {Proceedings of the Network and Distributed System Security Symposium},
  year          = {2010}
}

@inproceedings{slowinska2011howard,
  author        = {Slowinska, Asia and Stancescu, Traian and Bos, Herbert},
  title         = {{Howard}: A Dynamic Excavator for Reverse Engineering Data Structures},
  booktitle     = {Proceedings of the Network and Distributed System Security Symposium},
  year          = {2011}
}

@article{zhang2026thompson,
  author        = {Zhang, Wen and Chen, Jinfu and Cai, Saihua and Wang, Kun and Liu, Yisong and Ding, Haotong},
  title         = {A Novel Seed Scheduling Scheme Using {Thompson} Sampling for Coverage-Guided Greybox Fuzzing},
  journal       = {Journal of Systems and Software},
  volume        = {236},
  pages         = {112794},
  year          = {2026}
}

@article{cui2026visiongui,
  author        = {Cui, Chenhui and Huang, Yinming and Huang, Rubing and Zhou, Ling and Wang, Rongcun and Li, Tao},
  title         = {A Novel Vision-Based Approach to Test Sequence Generation for Mobile {GUI} Testing},
  journal       = {IEEE Transactions on Reliability},
  volume        = {75},
  pages         = {664--678},
  year          = {2026}
}

@misc{microsoftopenspecsrpc,
  title         = {{Microsoft Open Specifications: RPC}},
  howpublished  = {\url{https://learn.microsoft.com/en-us/openspecs/windows_protocols/ms-doco/711ba788-aa2d-432b-a863-957137b3d6d4}},
  note          = {Accessed in 2026}
}

@misc{microsoftsymbolserver,
  title         = {{Microsoft Public Symbol Server}},
  howpublished  = {\url{https://learn.microsoft.com/en-us/windows-hardware/drivers/debugger/microsoft-public-symbols}},
  note          = {Accessed in 2026}
}

@misc{microsoftpublicprivatesymbols,
  title         = {{Public and Private Symbols}},
  howpublished  = {\url{https://learn.microsoft.com/en-us/windows-hardware/drivers/debugger/public-and-private-symbols}},
  note          = {Accessed in 2026}
}

@misc{microsoftwin32api,
  title         = {{Windows API Reference}},
  howpublished  = {\url{https://learn.microsoft.com/en-us/windows/win32/api/}},
  note          = {Accessed in 2026}
}

@misc{microsoftmidl,
  title         = {{Microsoft Interface Definition Language}},
  howpublished  = {\url{https://learn.microsoft.com/en-us/windows/win32/midl/midl-start-page}},
  note          = {Accessed in 2026}
}

@misc{microsoftrpcndr,
  title         = {{RPC NDR Engine}},
  howpublished  = {\url{https://learn.microsoft.com/en-us/windows/win32/rpc/rpc-ndr-engine}},
  note          = {Accessed in 2026}
}

@article{liu2024lost,
  author        = {Liu, Nelson F. and Lin, Kevin and Hewitt, John and Paranjape, Ashwin and Bevilacqua, Michele and Petroni, Fabio and Liang, Percy},
  title         = {{Lost in the Middle: How Language Models Use Long Contexts}},
  journal       = {Transactions of the Association for Computational Linguistics},
  volume        = {12},
  pages         = {157--173},
  year          = {2024}
}

@inproceedings{modarressi2025nolima,
  author        = {Modarressi, Ali and Deilamsalehy, Hanieh and Dernoncourt, Franck and Bui, Trung and Rossi, Ryan A. and Yoon, Seunghyun and Schuetze, Hinrich},
  title         = {{NoLiMa: Long-Context Evaluation Beyond Literal Matching}},
  booktitle     = {Proceedings of the 42nd International Conference on Machine Learning},
  pages         = {44554--44570},
  year          = {2025}
}

@inproceedings{du2025context,
  author        = {Du, Yufeng and Tian, Minyang and Ronanki, Srikanth and Rongali, Subendhu and Bodapati, Sravan Babu and Galstyan, Aram and Wells, Azton and Schwartz, Roy and Huerta, Eliu A and Peng, Hao},
  title         = {{Context Length Alone Hurts LLM Performance Despite Perfect Retrieval}},
  booktitle     = {Proceedings of the 2025 Conference on Empirical Methods in Natural Language Processing: Findings of the Association for Computational Linguistics},
  pages         = {23281--23298},
  year          = {2025}
}

@inproceedings{wang2024clap,
  author        = {Wang, Hao and Gao, Zeyu and Zhang, Chao and Sha, Zihan and Sun, Mingyang and Zhou, Yuchen and Zhu, Wenyu and Sun, Wenju and Qiu, Han and Xiao, Xi},
  title         = {{CLAP: Learning Transferable Binary Code Representations with Natural Language Supervision}},
  booktitle     = {Proceedings of the 33rd ACM SIGSOFT International Symposium on Software Testing and Analysis},
  pages         = {503--515},
  year          = {2024}
}

@inproceedings{cummins2025llmcompiler,
  author        = {Cummins, Chris and Seeker, Volker and Grubisic, Dejan and Roziere, Baptiste and Gehring, Jonas and Synnaeve, Gabriel and Leather, Hugh},
  title         = {{LLM Compiler: Foundation Language Models for Compiler Optimization}},
  booktitle     = {Proceedings of the 34th ACM SIGPLAN International Conference on Compiler Construction},
  pages         = {141--153},
  year          = {2025}
}

@inproceedings{gao2026ink,
  author        = {Gao, Muhan and Chen, Zih-Ching and Huang, Kuan-Hao},
  title         = {{The First Drop of Ink: Nonlinear Impact of Distracting Information in Long-Context Reasoning}},
  booktitle     = {Proceedings of the 43rd International Conference on Machine Learning},
  pages         = {33019--33038},
  year          = {2026}
}

@inproceedings{zeng2026loca,
  author        = {Zeng, Weihao and Huang, Yuzhen and He, Junxian},
  title         = {{LOCA-bench: Benchmarking Language Agents Under Controllable and Extreme Context Growth}},
  booktitle     = {Proceedings of the 43rd International Conference on Machine Learning},
  pages         = {153347--153363},
  year          = {2026}
}

@misc{deepseekharness,
  title         = {{DeepSeek Harness}},
  howpublished  = {\url{https://github.com/deepseek-ai/deepseek-harness}},
  note          = {Accessed in 2026}
}

\end{document}